\documentclass[AMA,STIX1COL]{WileyNJD-v2}
\usepackage{subcaption}
\graphicspath{ {figs/} }

\articletype{Research Article}%

\received{26 April 2016}
\revised{6 June 2016}
\accepted{6 June 2016}

\begin{document}


\title{Honesty Based Democratic Scheme to Improve Community Cooperation for IoT Based Vehicular Delay Tolerant Networks}

\author[1,2]{Ghani ur Rehman}

\author[1]{Anwar Ghani*}

\author[2]{Muhammad Zubair}

\author[3]{Shahbaz Ahmad Khan Ghayyure}

\author[1]{Shad Muhammad}



\authormark{Ghani \textsc{et al}}

\address[1]{\orgdiv{Department of Computer Science \& Software Engineering}, \orgname{International Islamic University}, \orgaddress{\state{Islamabad 44000}, \country{Pakistan}}}

\address[2]{\orgdiv{Faculty of Computer Science \& Bioinformatics}, \orgname{Khushal Khan Khattak University}, \orgaddress{\state{Karak 27000}, \country{Pakistan}}}





\corres{*Dr. Anwar Ghani, Department of Computer Science \& Software Engineering, International Islamic University Islamabad, \email{anwar.ghani@iiu.edu.pk}}

\presentaddress{Department of Computer Science \& Software Engineering, International Islamic University Islamabad}

\abstract[Summary]{Many Internet of things (IoT) applications have been developed and implemented on unreliable wireless networks like the Delay tolerant network (DTN), however, efficient data transfer in DTN is still an important issue for the IoT applications. One of the application areas of DTN is Vehicular Delay Tolerant Network (VDTN) where the network faces communication disruption due to lack of end-to-end relay route. It is challenging as some of the nodes show selfish behavior to preserve their resources like memory, and energy level and become non-cooperative. In this article, an Honesty based Democratic Scheme (HBDS) is introduced where vehicles with higher honesty level are elected as heads -- during the process. Vehicles involved in the process would maximize their rewards (reputation) through active participation in the network activities whereas nodes with non-cooperative selfish behavior are punished. The honesty level of the heads is analyzed using Vickrey, Clarke, and Groves (VCG) model. The mathematical model and algorithms developed in the proposed HBDS technique are simulated using the VDTNSim framework to evaluate their efficiency. The performance results show that the proposed scheme dominates current schemes in terms of packet delivery probability, packet delivery delay, number of packets drop, and overhead ratio.}


\jnlcitation{\cname{%
\author{G. Rahman}, 
\author{A. Ghani}, 
\author{M. Zubair}, 
\author{S.A.Ghayyure}, and 
\author{S. Muhammad}} (\cyear{2020}), 
\ctitle{Honesty Based Democratic Scheme to Improve Community Cooperation for Internet of Things Using VDTN}, \cjournal{Journal name XYz, publisher}, \cvol{2020;00:1--16}
}

\maketitle

\footnotetext{\textbf{Abbreviations:} Delay Tolerant Networks, Selfish behavior, Honesty, incentive schemes, Internet of Things, Trustworthiness, cooperative communication}

\section{Introduction}
\label{intro}
Normally the traditional internet is the connectivity of homogeneous devices like computers but there are some emerging paradigm where 
Heterogeneous devices like tablets, computers, and smartphones are connected for some specific purposes in a wireless network like cloud and grid. One of the similar vision paradigms is the Internet of Things (IoT)~\cite{p00}. In IoT, the connected devices are heterogeneous and can be used for sensing, object identification, tracking of vehicles, privacy and traffic control and many other areas. In such scenarios, the underlying network is very important to consider. The IoT applications may suffer from low connectivity, mobility of nodes, and interruptions of links in the
urban environment. Such unstructured networks in the urban areas are termed as DTN. DTN ~\cite{p01} has nodes that cooperate to forward messages to its connected node. In the DTN the mobile nodes are intermediate and help in data transfer in the network. There are also connectivity issues in the DTN, so it used the store and forward routing method~\cite{p02,p03}. The node in the DTN stores the message until it finds another node to forward the message. DTN packet forwarding depends on the mobile node performance. Sometimes the mobile nodes are not willing to forward messages to other nodes in the network~\cite{p04}. It can be useful in Interplanetary, the
military, and (VDTN).

The paper is about the behavior of node (vehicles) in the vehicular Delay tolerant (VDTNs)~\cite{p05}, where the connected vehicles communicate wirelessly to each other. Many of the applications of the VDTN are traffic congestion notification, updates on the weather, emails and many others. The traditional routing protocols DSR and AODV may not work properly due to broken connectivity. The store and carry approach of these traditional routing has some cooperation for forwarding the data to its destination irrespective of the place. 
The messages are replicated by the DTN protocols to ensure the end to end connectivity issue. The selfish nodes can degrade the performance of the DTN network by showing misbehavior towards other nodes in a network. Some of the research has shown that the misbehaved nodes can badly damage the performance of the network~\cite{p06,p07},~\cite{p07a}. Other research has shown that the misbehavior is due to selfish nodes in the network. The selfish nodes save its resources such as (storage space, CPU time, and energy, etc) in order to be chosen~\cite{p08}. This clearly indicates that selfishness is the real issue in the DTN and needs to be addressed in a more sophisticated way. The two types of nodes in the network is normal and abnormal nodes~\cite{p09}. A normal type of nodes in the network actively participates in the network simulation and show its cooperation with its neighbor nodes. Abnormal nodes are those nodes which are not taking part in the network activities and can degrade the performance of the network. This selfish behavior of node is shown to save its resources like energy-saving, bandwidth, manipulation and other social behaviors~\cite{p10}. 

For the simulation of the selfish nodes in the network, many different mechanisms have been proposed. In the incentive and punishment-based scheme~\cite{p11}, nodes in the network get incentives based on cooperation and its weight. In multi-heads clustering scheme~\cite{p12}, the incentive to the nodes is rewarded only on its weight. The drawbacks with these two approaches are that, (i) the contacts made are only for short duration in IPS Scheme (ii) there is chances of weight tie problem in multi-leader election scheme. The proposed scheme omits the selfish behavior of the nodes in different way as compared to these techniques like nodes nomination criteria for the democratic system is based on some honesty parameters. It is observed that when the nodes in the network have long and frequent conversations, higher level of honesty is expected. The following are the key contribution in the proposed scheme in VDTN;
\begin{itemize}
\item Analyzing node behavior in VDTN.
\item To investigate the influence of the activities of the node on network performance and to develop a methodology in which node with selfish behavior are stimulated and encouraged to cooperate. 
\item To boost the overall performance of the system on the basis of node honesty with different parameters such as frequency of interaction, mutuality and centrality and community of interest to elect heads e.g. Community Head, Auxiliary Community Head and Incentive Head.
\item To design a Watchdog system that properly check the behavior of nodes in a community.
\item A Comparative review is conducted to compare the performance of the proposed mechanisms with all the other incentive-based techniques. In addition, the proposed solution has increased network performance in terms of packet delivery probability, packet dropping ratio, overhead and delivery delay.
\end{itemize}
Rest of the paper is organized as: Section~\ref{sec:relwrk} discusses related work, in Section~\ref{sec:prop}, HBDS detailed design is presented, it also covers system model. The performance results of HDBS is presented in Section~\ref{sec:per}. The paper is finally concluded in Section~\ref{sec:con}.    
\section{Related Works}
\label{sec:relwrk}
\noindent IoT has greatly influenced the smart cities and helps it in making a sustainable environment in it. The major objectives of smart and sustainable societies are to provide livability, renewal, and a sustainable environment for human life~\cite{p12a}. Smart and sustainable cities are connected by heterogeneous devices like computers, traffic lights, roadside units, healthcare units, and smart homes. One of the important aspects of smart and connected cities~\cite{p12b},~\cite{p12b1} is to make societies sustainable. The connected objects in such types of societies are to provide a resource-efficient environment, ensure safety, healthy living-hood, and provide a smart transport system~\cite{p12c,p12d}. The IoT applications in the smart transport system depend on the VDTN paradigm of networking~\cite{p12e}. In VDTN, the vehicles are connected with roadside units, lights, and other vehicles. For IoT applications to work in VDTN efficiently the connected objects need to be cooperative and forward data within a network~\cite{p12f}. However, some of the nodes in the VDTN have selfish behavior. Many researchers have addressed the issue of selfishness in it. The selfish behavior of the nodes degrades the performance of the network~\cite{p13},~\cite{p13a}. Incentive-based mechanisms encourage the selfish nodes to simulate and share its resources for cooperation in the network~\cite{p14}. Incentive schemes are credit-based, reputation-based, game-theory and barter systems. All the cooperative nodes are rewarded for showing cooperation by the credit-based approach in the network. This reward will be used by the nodes in the network for their purpose later on in the election process. 

Jiang et al.~\cite{p15} have designed a secure credit-based incentive scheme (SCIS)to encourage the selfishness nodes for cooperation in opportunistic networks. The selfish nodes are given some sort of incentive in the form of virtual credit for their cooperation. Ning et al.~\cite{p16} proposed a Copy Adjustable Incentive Scheme(CAIS) is a type of mechanism where all the nodes in the community are divided based on the social interaction in the network. In this scheme, two types of incentive namely social and non-social credits are given to the nodes. Park et al.~\cite{p17} have proposed a bitcoin-based secure incentive scheme to motivate nodes for cooperation in VDTN. Bitcoin is a very popular currency and digital payment based some techniques of cryptography. Jiang et al.~\cite{p18} proposed a credit-based congestion-aware incentive scheme (CBCAIS) to handle the issue of selfishness in DTN. Nodes are given credit for their cooperation. The selfish nodes are punished for not showing cooperation in the network. Jedari et al.~\cite{p19} proposed a game-theoretic incentive scheme for social-aware routing in selfish mobile social networks (GISSO) to stimulate the selfish nodes for cooperation. 

The reputation-based incentives scheme calculates the nodes cooperation in the network. Cooperative nodes in the network have higher values than the selfish and non cooperative nodes. Rehman et al.~\cite{p11} proposed an incentive and punishment scheme for to discuss the issue of selfishness. In this scheme on one hand nodes are given an incentive in the form of a reputation for their cooperation within the community. On the other hands, nodes are punished for showing selfish behavior repeatedly in the form of expulsion from the community. The problem of selfishness in Vehicle Delay Tolerant Network is presented in Dias et al.~\cite{p20}. In this approach, the score for a successful transfer of data is calculated. If packets are received successfully, the score will be increased; otherwise, a certain number will decrease the score. Mantas et al.~\cite{p21} proposed a reputation-based scheme to handled the issue of selfishness. In this scheme monitor nodes are used to properly monitor the behaviors of the nodes in the network and nodes are encouraged for cooperation. Jedari et al.~\cite{p22} proposed So-Watch (Social Watchdog system) uses a reputation scheme that identifies the number of selfish nodes and the level of selfishness. Kou et al.~\cite{p23} proposed a strict reward and punishment model to tackled the issue of selfishness. Nodes are stimulated for taking part in the routing process. Sharma et al.~\cite{p24} proposed a reputation-based scheme to detect selfish and malicious nodes and also stimulates such nodes for cooperation in Delay tolerant networks. Cai et al.~\cite{p25} have presented an efficient incentive-compatible routing protocol (ICRP) for DTNs. In this scheme nodes are rewarded incentives in the shape of a reputation for showing cooperation. 

In barter system, the nodes in contact share same amount of information which is also called tit-for-tat scheme. Liu et al.~\cite{p26} proposed that message can be forwarded between the communities and the nodes based on the barter-based mechanism. The nodes and the community in the network share the messages.  The messages are shared between node and community. Zhou et al.~\cite{p27} that Tit-for-Tit scheme for the message forwarding in the network. In this scheme a reward is given to the cooperative nodes for forwarding of data to the other nodes. Buttyan et al.~\cite{p28} proposed a barter-based approach to deal with issue of selfishness.  

Some other approaches such as game-theoretic, social and trust-based have also been proposed to deal with the issue of selfishness. SSAR~\cite{p29} is a social-based scheme that tackled the problem of selfishness in a network. In this approach, a node having a strong social relationship in a community can take part in data forwarding for other nodes. Lui et al.~\cite{p30} proposed an incentive scheme to share information that is related to a road accident, road construction and delay in traffic, etc. Fawaz et al.~\cite{p31} proposed Unmanned Aerial Vehicles (UAVs) scheme to encourage vehicles for cooperation. Li et al.~\cite{p32} proposed a scheme which also stimulates vehicles for cooperation in a vehicle to vehicle networks. Socievole et al.~\cite{p33} proposed a social approach to encourage selfish and misbehavior nodes for cooperation. Umar et al.~\cite{p34} proposed a game-theoretic scheme that motivates the nodes for cooperation to reduces the load on nodes. Vamsi et al~\cite{p35} proposed a trust-aware scheme to stimulate the nodes to take part in the routing process. Kumar et al.~\cite{p36} has proposed a dynamic trust-based intrusion detection scheme track and isolate the selfish nodes from the network. Pal et al.~\cite{p37} proposed a trust-based approach for ensuring the routing process. Dhurandher et al.~\cite{p38} proposed a trust-based scheme that motivates the nodes for cooperation in the network.
\section{Proposed Scheme HBDS}
\label{sec:prop}
\begin{figure}[htbp]
\centerline{\includegraphics[scale=.7]{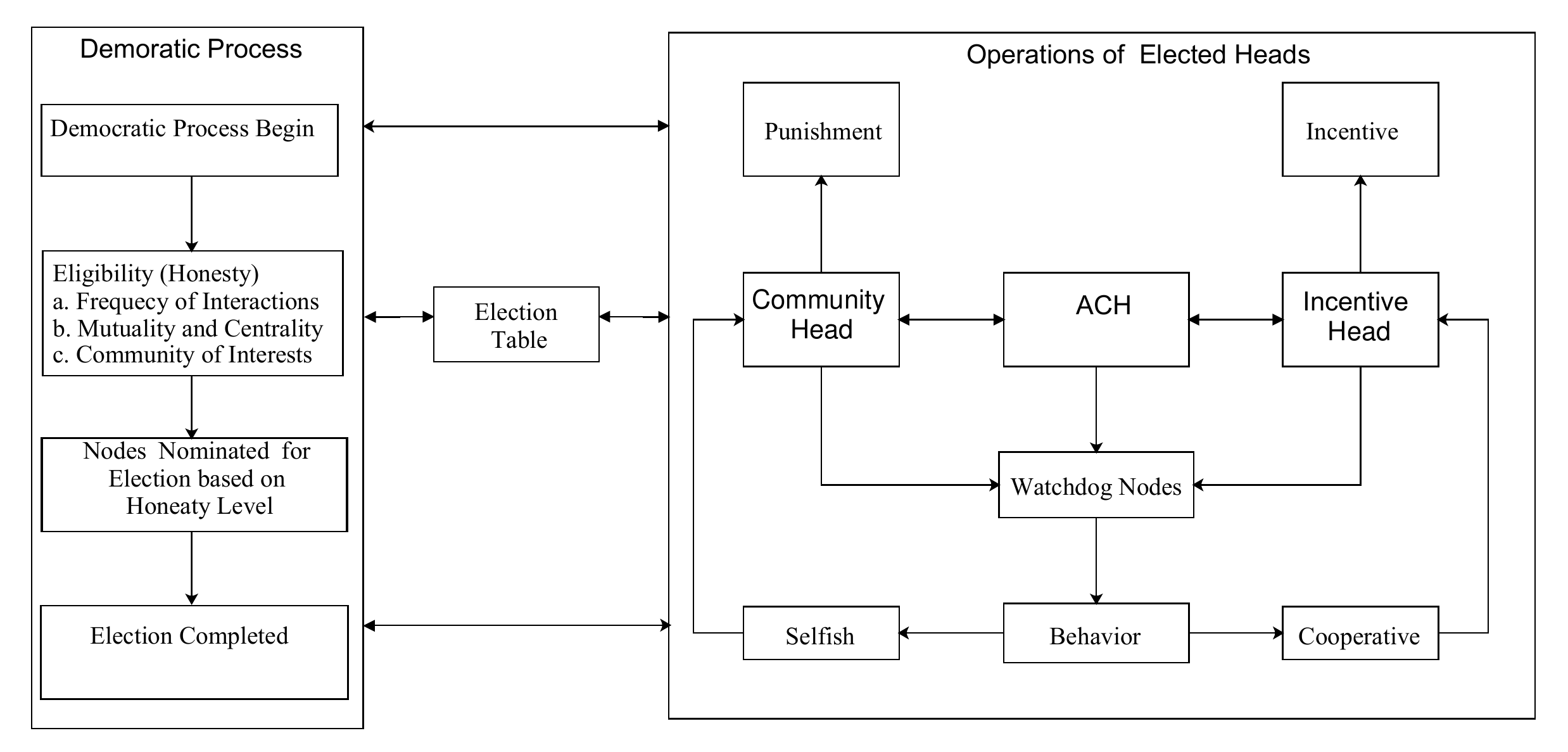}}
\caption{Overall Structure of Proposed Scheme}
\label{fig}
\end{figure}

The prime focus of the proposed scheme HBDS is to investigate the involvement of node in DTN network. The nodes in the network are responsible for message forwarding and watching over its fellow nodes. These two activities are main responsibilities of the nodes in the democratic process. Nodes are encouraged to participate in the network and coordinate to make better performance of the network. In a community based network the node watches the fellow node behavior for sending and receiving messages. This gives controls to the node over message forwarding in a network. The proposed mechanism suggests incentives to the nodes in the form of reputation to motivate the nodes for simulation and to improve the performance of the network. The proposed mechanism will show us the involvement of the nodes in the network and its contribution towards network performance. The contribution of the node is from message forwarding to the watching of the fellow nodes for the message forwarding. The proposed mechanism makes incentive to the nodes for a) involvement of the node in the Democratic process b) Messages forwarding to its fellow nodes c) Watching the behavior of the other nodes. These will be explained in later sections. The nodes with selfish behavior in the network are simulated and encouraged to participate in the democratic process to cooperate in the networks. The node with persistent selfish behavior are punished and may be expelled from the community and this message will be broadcasted in the community. The registration of the nodes in the VDTN will be decided by centralized authority. Registration process depends on three parameters, per-node budget, payment to the relay node, and the payment made to the watchdog. The working of the proposed scheme presented in the figure~\ref{fig}.  

\subsection{Incentive for Participation in Democratic Process}
The proposed HBDS scheme forwards the data to other nodes in the community. The nodes in the network start the democratic process by participating in the election. Some of the topics like community formation and its maintenance can nor be covered in this paper and therefore it is not discussed here.

\subsubsection{Democratic Process}

The democratic process periodically manages the community. The eligibility criterion of the democratic process is trust attribute (TA) honesty of the node. It plays an important role in the assessment of the trust developed among the nodes in the network. Without honesty the trust among the nodes will not be genuine and strong relationships cannot exist as in the sociology concept. Trust attribute (TA) shows the trustiness of the vehicle with the trustor vehicle~\cite{p39}. The honesty property of the trust is chosen because dishonest nodes can make delays in the network and can obstruct the overall operations of the network delivery. Honest nodes will be nominated for the election process. The three factors for the node honesty calculation are (a) Frequency of interaction (b) Mutuality and Centrality and (c) Community of interest. The node that scores the highest number of votes for honesty will be elected as community head $CH$. Second, highest sorcerer will be elected as auxiliary community head $ACH$ and the third-highest scorer will become Incentive head $IH$. To bring stability to the community, the $ACH$ can become $CH$ if the community head moves away from the community and is not part of the community anymore. Heads in the community are responsible for assigning unique community-id. After every democratic process, the $CH$ uses MAC-id to generate community-ids to the nodes in the community. The unique community-ids are stored and maintained by the succeeding $ACH$. The incentive head $IH$ makes payment to the cooperative nodes in the form of incentives. The watchdog nodes are used to monitors the  behavior of the nodes and are selected by the heads in the network. Honesty calculation is a lengthy process. It combines past information with the new information to predict the honesty level of the node. The honesty level depends on the frequency of conversations and the length of the conversation in the connected nodes. The cooperation of a node can also be judged by the balanced of interactions among the nodes. $H_{xy}^{a}(t)$ provides the evaluation of honest to object $y$ by object $x$ at time $t$, where $a$ represents the attributes. The calculation of node honesty is explained as follow:

\paragraph{Frequency of Interaction:} Let us consider a conversations between nodes $C=\{c_1,c_2,,c_3,……,c_n\}$, the conversation between trustee nodes over time t. On the basis of this, the degree of honesty between two nodes can be calculated in equation \ref{eq:01}.
\begin{center}
\begin{equation}
\label{eq:01} 
Honesty_{xy }^{TC} (t)=\sum_{i=1}^{j} \frac {\mid cl_{i}\mid} {\mid tcl_{i}\mid} E(cl_{i})
\end{equation}
\end{center}
where $j$ is the number of contacts made, means the frequency of interactions of a node with each other, $cl_{i}$ is the length of the $i^{th}$ successful communication. $tcl_{i}$ is the overall length of communication and $E(cl_{i})$  will measure the usefulness of the conversation and is called the entropy function.

\paragraph{Mutuality and Centrality:} Common friends of two nodes show the honesty of the trustor and trustee. More common friends between the nodes mean higher reliability of communication between them. We can calculate the mutuality and centrality (credibility) of the trustee node and is given in equation \ref{eq:02}.
\begin{center}
\begin{equation}
\label{eq:02} 
Honesty_{xy }^{cen} (t)=\frac{\mid\mid M_{xy}\mid\mid}{\mid\mid N_{x}\mid\mid}
\end{equation}
\end{center}

where $M_{xy}$ is the number of common friends between two nodes and $N_{x}$ is the trustee friends.

\paragraph{Community of Interest:} We know that both the trustor and the trustee can be members of more than one community. One of the important factors in honesty is to see how much interest a node takes in a community. It also shows the common capabilities of nodes in a community. Here $M_xy$ shows the common communities of the trustor and the trustee nodes and $N_xy$ shows the communities of which trustee is a member is shown in equation \ref{eq:03}.
\begin{center}
\begin{equation}
\label{eq:03} 
Honesty_{xy }^{coi} (t)=\frac{\mid\mid M_{xy}^{coi}\mid\mid}{\mid\mid N_{x}^{coi}\mid\mid}
\end{equation}
\end{center}
Once the values are calculated for all three parameters, then the final honesty degree of a trustee is given in equation~\ref{eq:04}.
\begin{center}
\begin{equation}
\label{eq:04}  
FH_{xy}(t)=\alpha_{1}Honesty_{xy}^{TC}(t)+\alpha_{2}Honesty_{xy }^{cen}(t)+\alpha_{3}Honesty_{xy }^{coi} (t)
\end{equation}
\end{center}
where $FH$ is the final honesty, the $(\alpha_{1},\alpha_{2},\alpha_{3})$ are selected randomly, where the total honesty is equal to 1. Algorithm~\ref{alg} has detail about the democratic process in the proposed scheme. 
\begin{algorithm} 
\caption{Democratic Process and Incentive algorithm} 
\label{alg} 
\begin{algorithmic}[1]
\Require Number of nodes $n$
 \Ensure  Democratic Process and Incentive to nodes participating in the Democratic process
   \For{$x = 1:n$}
		\State{Compute and broadcast $FH_{x}$}
	    \EndFor
    \ForAll{$k\in n$} 
    \State{Nominate node $max(FH_{k})$ for election}
	\EndFor
	\State{Votes are counted and nodes such as $CH$, $ACH$ and $IH$ elected in democratic Process}  
	\State{$Pay=\sum_{k\in n}(Vt
			_{x}(FH,k)).Fb.\beta_{x}$ and $Cost_{x}=\frac{1}{FH_{x}}\ast(FH_{y}-FH_{x})\sum_{k\in n}(Vt_{x}(FH,k)) Fb.\beta_{x}$}
	\State{$CH$ new reputation, $Rep_{CH}=Rep_{CH}+Pay_{CH}-Cost_{CH} $}
	\ForAll{$k$, $k$ is not CH in community}
	\State{new reputation $Rep_{k}=Rep_{k}+Pay_{k}$}
	\EndFor
	\ForAll{$k\in n$}
	\State{broadcast $CH_{ack}$=$Vt_{CH}(k)\parallel Pay_{k}\parallel Rep_{k}$}
   \EndFor
	\State{Update Election Table}
\end{algorithmic}
\end{algorithm}
	
In this algorithm, honesty of all nodes are calculated and then broadcasted in the community. The nodes are nominated for democratic process on the basis of honesty and then incentive is given to all the nodes participating in the democratic process. The running complexity of the algorithm is $O(nk)$.

A node in the community with a higher honesty level is considered as a candidate for the democratic process of election. The node sometime can declare itself as honest and report false information. The false report will be either declaring itself as dishonest or over an honest node. To escape the chances of becoming community head the node declares itself under honest and it may declare itself over honest to get some incentives to become community head. The trust behavior of the nodes in a community will be evaluated by VCG model. To know about the false information reported by the nodes, VCG model is used. 

\subsubsection{Vickrey, Clarke and Groves (VCG) Scheme}

Vickrey Clarke and Groves is an important and valuable method of game theory. It predicts the nodes behavior in the network and motivates the dishonest nodes to participate in the network~\cite{p40}. It was previously assumed that energy level and weight of the node as internal and private information of the node. So the truth-telling behavior of the VCG was used to predict the behavior of the node~\cite{p41}. In our proposed model, some of the parameters of the nodes declare the honesty of the respective node. The honesty level of the node is considered as internal and private information. The mobility parameters of the node show the honesty level of the node in VDTN. The reputation of the node reaches to certain level after every democratic process. The reputation of the node s calculated on the basis of cooperation of the node in the network. The payment procedure detail is given in Algorithm~\ref{alg1}.

\begin{algorithm}[h] 
\caption{Responsibilities of Elected Heads and Incentive for packet forwarding algorithm} 
\label{alg1} 
\begin{algorithmic}[1]
	\Require Number of nodes $n$
  \Ensure Operations of Elected Heads\& Incentive to nodes for Packet Forwarding
  \For{$x = 1:n$}
  	\State{for each relay in community}
	\State{Elected Heads collectively assign Watchdog nodes WN1, WN2 and WN3}
	\State{$WN_{i}, \{ i=1, 2, 3\}$, wait for time ‘t’}
	\If{The same packet overheard after time ‘t’}
	\State{beh = Coop;}
	\State{send $t(report) = 'Coop'$}
	\Else
	\State{beh = Self;}
	\State{send $t(report) = 'Self'$}
	\EndIf	
	\State{$CH$ compute $IA$}
	\State{$IA_{x}=\frac {R_{x}}{\Sigma^{3}_{x=1} R_{x}}$}
	\State{$CH$ Calculate the behavior of all nodes $WN_{i}$}
	\If{$x_{coop}>x_{self}$}
	\State{grant $Pay_{r}=Rep_{r}+F_{pay}$}
	\Else
	\State{grant $Pay_{r}=Rep_{r}-F_{pay}$}
	\EndIf
	\If{$t_{report}(WN_{i})=t_{final}$}
	\State{grant $Pay_{WN_{i}}=Rep_{WN_{i}}+Pay(WN)$}
	\Else 
	\State{grant $Pay_{WN_{i}}=Rep_{WN_{i}}-Pay(WN)$}	
	\State {update $R_{Table}$}	
	\EndIf
	\EndFor
\end{algorithmic}
\end{algorithm}

In this algorithm watchdog nodes are collectively selected by the three heads. The watchdog nodes are responsible for monitoring the behavior of all nodes in the network. The importance aspect of all watchdog nodes are calculated. The overall operations and responsibilities of the elected heads and payment detail is also given in the  algorithm. The complexity of the algorithm is $O(n)$. 

\subsubsection{Payment after Democratic Process using VCG Model}

The game has $n$ number of players. We consider each node to be a community player. The democratic process starts with nodes disclosing their honesty level. Democratic election process is used for nodes election as heads and declared others as participants. The heads elected through the democratic process and other participants will receive payments in the form of reputation. Nodes in the network are always in search to increase its reputation value $R$. the higher the value of $R$ the more services they receive from the network. Each node in the network has $RTable$ to maintain its reputation. The node updates its $Rtable$ when the node gets updates from neighbor node.

\paragraph{Payment to Elected head $(CH)$ of Community}

Each node gets payment after the democratic process ends. The $CH$ gets payment on the basis of a number of votes it receives from the participants nodes. the nodes participating in the democratic process also get payments for voting, which denote the cost of the elected $CH$. $H_{1}, H_{2}, ...., H_{n}$, where $n$ is all number of nodes, represents the cost vector (nodes honesty). Difference in the value of payment to the node and payment received is the gain of $CH$ is given in equation \ref{eq:05}.
\begin{center}
\begin{equation}
\label{eq:05}
 Pay=\sum_{k\in n}(Vt_{x}(FH,k)).Fb.\beta_{x}
\end{equation}
\end{center}
where $Vt_{x}(FH,k)$ provides certain values in the democratic process (value is $1$ if $k$ node votes for $x$, otherwise 0 will be produced. $CH$ also assigns a fixed budget $FB$ to every node participating in the democratic process. This fixed budget is known to all the nodes in the network. $\beta_{x}$ is a payment of node shown below in equation~\ref{eq:06}.
\begin{center}
\begin{equation}
\label{eq:06}
\beta_{x} = FH_{x}+\frac{1}{\sum_{k\in n}Vt_{x} (FH,k)}\times \sum_{l\in n}(FH_{y})\times{\sum_{k\in n}Vt_{y}}(FH|FH_{x}=\infty,k)-{\sum_{y\in n}}(FH_{y}){\sum_{k\in n}}(Vt_{y}(FH,K))
\end{equation}
\end{center}

\paragraph{Payment to Community Members} 
The total cost of the nodes is divided among them based on the reputation of the node. The $CH$ announces the payment to the nodes through $CH_{ack}$ message. All the messages once received are authenticated by the standard message authentication mechanism. An update is made to the RTable in each node. The fixed payment $FB$ is used to distribute the total cost among the connected nodes who have participated in the voting process. We can calculate the cost function $C_{x}$ determined by $CH_{x}$ and is given in equation \ref{eq:07}.
\begin{center}
\begin{equation}
\label{eq:07}
 Cost_{x}=\frac{1}{FH_{x}}\ast(FH_{y}-FH_{x})\sum_{k\in n}(Vt_{x}(FH,k)) Fb.\beta_{x}
\end{equation}
\end{center}
where $FH_{x}$ and $FH_{y}$ represent the highest and second-highest nodes honesty of the nodes participating in a democratic process. As noticed below, the elected heads calculates their reputation by deducting the overall cost form their own payment to calculate their own reputation and is shown in equation \ref{eq:08}.
\begin{center}
\begin{equation}
 \label{eq:08}   
 Rep_{x}=Pay_{x}-Cost_{x}
\end{equation}
\end{center}
Each node gets a share of the total cost based on the reputation of the node. $CH_{ack}$ message is used to announce the payment to the nodes in the network. These messages are verified by the standard authentication messages. After each announcement the node updates its $RTable$.

\subsection{Packet Forwarding Payment}

The gateway nodes(GW) and the community head (CH) cannot be used as a relay node. This restriction means that these nodes cannot be used to forward packets. This is a negative and selfish behavior of the node and can degrade the performance of the network. Selfish nodes can make disconnections and drop packets, this selfish behavior has a negative impact on the performance of the network. To stimulate the nodes for participation in the network, some incentives have been awarded in the form of reputation. To make the payment system effective, a monitoring system is used to monitors the relay nodes. Table~\ref{table1} has the notations used in the proposed scheme.
\begin{table}
\center
\caption{Summary of notations and symbols}
\label{table1}
\begin{tabular}{ll}
\toprule
 Notation  &            Representation of the symbol\\
\midrule
$x$ & $CH$ is the Community Head that is elected in Democratic Process\\
$cl_{i}$ & Lenght of $i^{th}$ successful communication\\
$tcl_{i}$ & Overall length of communication\\
$E(cl_{i}$ & Entropy Function\\
$Rep_{x}$ & Node $m$ reputation\\
$Pay_{x}$ & Node $x$ payment\\
$Vt_{x}$ & Node $k$ which voted for $x$ \\ 
$\beta_{x}$ & Per member node cost\\
$H_{x}$ & Honesty level of node $x$\\
$IA_{x}$ & Importance aspect of node $x$\\
$M_{xy}$ & Number of common friends\\
$N_{x}$ & Trustee Friends\\
$coi$ & Community of interest\\
$WN$ & Watchdog Node\\
\bottomrule
 \end{tabular}
 \end{table} 

\subsubsection{Incentive for Relay Nodes} 

The $IH$ makes payment to the nodes for each data forwarding in our proposed scheme. It pays fixed payment $(F_{pay})$ in the form of incentives and is managed by incentive head. The $F_{pay}$ is allocated to the relay node on the basis of its cooperative behavior. To monitor this a watchdog system is introduced to monitors the behavior of all the relay nodes and then take the behavioral decisions. 

\paragraph{Cumulative Trust of Watchdog Nodes}

In Proposed scheme, we have three watchdog nodes in a relay. One of them is auxiliary community Head $ACH$, the second one is a predecessor to the $ACH$ and the third one is selected in round-robin as member node. The nodes in the relay create hash~\cite{p42}, the function of the hash is to keep the packet genuine and prevent the forwarding node from changing the packet. Every packets when reach to its destination nodes, it is verified with hash function value. If the value of the has function is matching, it means the packet is in steady-state. If the hash value is not matching then the forwarder node is declared as a selfish node, and it receives negative payment. The nodes in the network keep a record of the recently forwarded packets in a buffer. Each packet has a certain expected time to be sent next. So we can calculate direct and indirect trust based on the node behavior. Direct trust is based on the interaction of the node with its neighbor nodes. There are three watchdog nodes in our proposed scheme. The watchdog function as monitor to observe the behavior of the neighbor nodes in the network. Any unusual behavior is reported and trust value is calculated using a hash function. A watchdog system constantly monitors the conversation of the node with its neighbor node. It calculates the total number of dropped and forwarded packets in a network. The mobility of the node creates difficulty in calculating the trust value. It is because the node while moving also discover new nodes and replaces an old neighbor. To overcome this situation, the second opinion is taken to assess the rigorous trust value. So indirect trust is calculated to improve the overall decision-making system. So each watchdog calculates both direct and indirect trust value for the total calculation of the trust value. The range value is [0,1] for each watchdog node. The initial value starts at 0.5. Any value greater than 0.5 is termed as a cooperative node and less than 0.5 is a selfish node.

The trust reported generated by the node is its individual opinion. It can be cooperative or selfish, based on the individual behavior of the node. The $CH$ calculates the total trust value from three trust reports of the watchdog nodes. If the total value of the trust is greater than the selfish behavior of the node, the nodes is declared as genuine and incentive in the form of reputation is made to it. In order to avoid contradiction in the results, the $CIA$ cumulative importance aspect rule is introduced to calculate the trust value from different nodes based on certain evidence.

\paragraph{Cumulative Important Aspect for Trust Calculation}

The $CIA$ can report any bias, if the participating nodes have any prior indulgence of declaring each other as cooperative nodes. It may also unjustifiably declare a node as selfish. So the $CIA$ makes this distinction of the selfish and cooperative node. The watchdog nodes important aspect is its honesty. The honesty of the watchdog node is directly linked to its reputation. The watchdog node $IA$ important aspect is equivalent to the reputation level over the cumulative reputation level of all the watchdog nodes in the community. $IA_{x}$ indicates the importance of the watchdog node and $R_{1}, R_{2}$ and $R_{3}$ are three relay-involved watchdog nodes and is given in equation \ref{eq:09}.

\begin{center}
\begin{equation}
\label{eq:09}
IA_{x}=\frac {R_{x}}{\Sigma^{3}_{x=1} R_{x}}                        
\end{equation}
\end{center}

If any $x$ node reports the behavior of the any $n$ node, the accurate judgment is considered to be equivalent to the importance aspect $(IA)$ of a node reporting the nodes behavior. Therefore, any $x$ node with an important aspect $IA_{x}$ states that node $x$ is cooperative and is given in equation~\ref{eq:10}.
\begin{center}
\begin{equation}
\label{eq:10}
M_{x}(Coop)=IA_{x}
\end{equation}
\end{center}
\begin{center}
\begin{equation}
\label{eq:11}
M_{x}(Self)=1-IA_{x}
\end{equation}
\end{center}

Moreover, If any $y$ node reported $k$ as selfish then
\begin{center}
\begin{equation}
\label{eq:12}
M_{y}(Self)=IA_{y}
\end{equation}
\end{center}
\begin{center}
\begin{equation}
\label{eq:13}
M_{y}(Coop)=1-IA_{y}
\end{equation}
\end{center}
The evidence from all the nodes will be used to calculate the final honesty level of the node to award payment. The $CIA$ rule decides the behavior of the node as selfish or cooperative. The $CIA$ rule state that even if two of the watchdog nodes declare the node to be selfish, but the total value of trust is less than the third watchdog node that claims the node to be cooperative. Then according to $CIA$ rule, the node will be acknowledged as a cooperative node. The final reputation is calculated from the awarded payment to the node and current reputation. This is announced by the $CH$ in the network to all.

\subsubsection{Watchdog Nodes Payment} 

Each trustworthy report submitted by the monitoring node gets payment for it. The $CH$ makes payment to the nodes on their trust value. The final value is the trustworthiness of the monitoring node. For $(Pay>0)$, shows that this node is trustworthiness as the final trust value is matching with trust report. For $(Pay<0)$, it shows that the final trust value is different from the trust value and it termed the node as misbehaving watchdog node. Some slight modification is made to deal with different features of the participating devices in the community. So the honesty of the node is considered as internal and private to the node. This is the basic eligibility criteria for the participation in the democratic process. The awarded value of the reputation is the real number of each node. This value alters as the truth-telling behavior of the nodes changes. 

\subsubsection{Carry-Forward Reputation}

The initial reputation of each node in the network is set to zero. The reputation of the node changes during the election process. The change made to the reputation value is updated and $CH$ node announces it after every update. The node status changes to $GW$ after every update from the $CH$ node. Every node shares its reputation table $RTable$ with community head $CH$ after each update. The data forwarded by the $GW$ node is checked for authentication by the $CH$ before it is broadcasted. This means that all the information about the reputation is managed and stored by the $CH$ in a community. Let us suppose that there are three community $a$, $b$ and $c$, and new node $N$ enters in the $b$ community. According to our proposed scheme community, $b$ knows about the reputation values of all the participated nodes in the $a$ and $c$ community. So the community $a$ and $b$ has all the information about the $N$ node even before it enters into $a$ and $b$ community. If $CH$ has no information about the new node, then this new node will be termed as a new node to the network. 

\subsection{Penalties for Selfish Nodes}

The community head stimulates the selfish nodes in the network for participation in the democratic process. It can also punish the selfish node with repeated misbehavior in three ways. The $CH$ gives zero incentive to the nodes if it behaves selfishly for the first time and encourages it to participate. The $CH$ gives negative payment to the node if it shows selfish behavior for the second time. The $CH$ node can punish the node with repeated selfish behavior and expel it from the network for some time. It is also noted that the node may join the network after some time and cooperate with the nodes, so such a node should pay the negative payment first. The penalty coefficient is used as an importance mechanism for feedback. This feature is used to measure the dishonesty level of the node, which has shown such behavior in the past. It is important to keep the honesty level at a certain level, so an exponential downgrading system is used. $\mid\mid C \mid\mid$ shows the total number of conversations at time $t$ and $\mid\mid U_{C}\mid\mid$ is ineffective conversations. The penalty is calculated in equation \ref{eq:14}:
\begin{center}
\begin{equation}
\label{eq:14}
H_{p}(t)=\frac{\mid\mid {C}\mid\mid-\mid\mid U_{C}\mid\mid}{\mid\mid C \mid\mid}e^{( \,{\frac{\mid\mid {C}\mid\mid-\mid\mid U_{C}\mid\mid}{\mid\mid C \mid\mid}}) \,}
\end{equation}
\end{center}
\section{Performance Evaluation}
\label{sec:per}
\noindent
In this section, a simulation tool VDTNSim was used to conduct the simulation performance of HBDS scheme~\cite{p43}. VDTNSim is the extended version of the Opportunistic Network Environment (ONE) Simulator~\cite{p44}. This makes it possible to simulate the VDTN architectural solution, which involves the store-carry-and forward relayed network just below the network layer. The Simulation setup, performance metrics, and results are shown in the next subsections.

\subsection{SIMULATION SETUP}
\noindent
The simulation takes place in three modules. The first module shows the evaluation of selfish nodes effects on performance metrics such as packet delivery probability, packet dropping ratio, overhead and delivery delay. Based on our setup the results for the HBDS protocol are obtained in the second module. In the third module, the results are contrasted and examined with two existing protocols. Selfish node differences are used as techniques for testing all three protocols. For 0\% selfish nodes, the network is assumed to be normal. The selfish nodes range from 0\% to 80\% in the network throughout the simulation process. SSAR is used as a basis for comparison. Table~\ref{table2} lists the parameters used in the simulation.
\begin{table}
\center
\caption{Simulation Parameters}
\label{table2}
\begin{tabular}{ll}
\toprule
 \textbf{Parameters}  &           \textbf{Values}\\
\midrule
Simulation Area	&  $4500\times 3500 m^2$\\
Number of nodes&  100\\
Transmission Range& 300m \\
Malicious Activity &0\% to 80\%\\ 
Comparison & Proposed Scheme compared with SimBet and SSAR    \\
Simulation Time &24 hrs\\ 
Number of Relay Nodes& 05 \\
Number of Terminal Nodes & 100\\
Terminal Nodes Buffer Capacity& 150 MB\\
Relay Node Buffer Capacity& 250 MB \\
Average Speed& $60 km/h$\\
Variation in Nodes& 20 to 100\\
Interval for Packets Generation& [20, 30] sec\\
Node Communication& IEEE 802.11b\\
Size of Packets& [50, 650 KB]\\
Packets TTL& 320 minutes\\
\bottomrule
\end{tabular}
\end{table} 
\subsection{METRICS} 
The simulation uses packet delivery probability, packet dropping ratio, delivery delay, and overhead ratio as metrics of performance within a network. The probability of packet delivery was computed as the percentage between the number of the individual packets forwarded to their intended destination and the number of individual packets created at source nodes. The packet delivery delay for the packet is calculated as the average amount of time packets have to transfer from source to destination. A nodes in the network dropped packets because there is lesser space in the buffer, TTL expiry is soon and selfish behavior of the node. The overhead ratio calculates a routing protocols bandwidth performance. It means it has to calculate extra bundles for necessary delivery that packet. HBDS is compared with incentive-based protocols: SimBet~\cite{p45} and Social selfishness Aware Routing (SSAR)~\cite{p29}. SimBet is used as a basis for comparison.

\subsection{RESULTS AND DISCUSSION} 

VCG method has been used to make payments to the nodes. The nodes participating in the democratic process are rewarded incentives to show collaboration and cooperation in a network. Such cooperative nodes can become Community Head, Auxiliary Community Head, and Incentive Head via the democratic process. Negative incentives are provided to selfish nodes for not exhibit the appropriate commitment. The Nodes are also punished in the form of removal from the network if they consistently exhibit selfish behavior.

\subsubsection{Impact of Selfish Nodes on Performance Metrics} 
\begin{figure}[ht]
\begin{subfigure}{.5\textwidth}
  \centering
  \includegraphics[scale =.43]{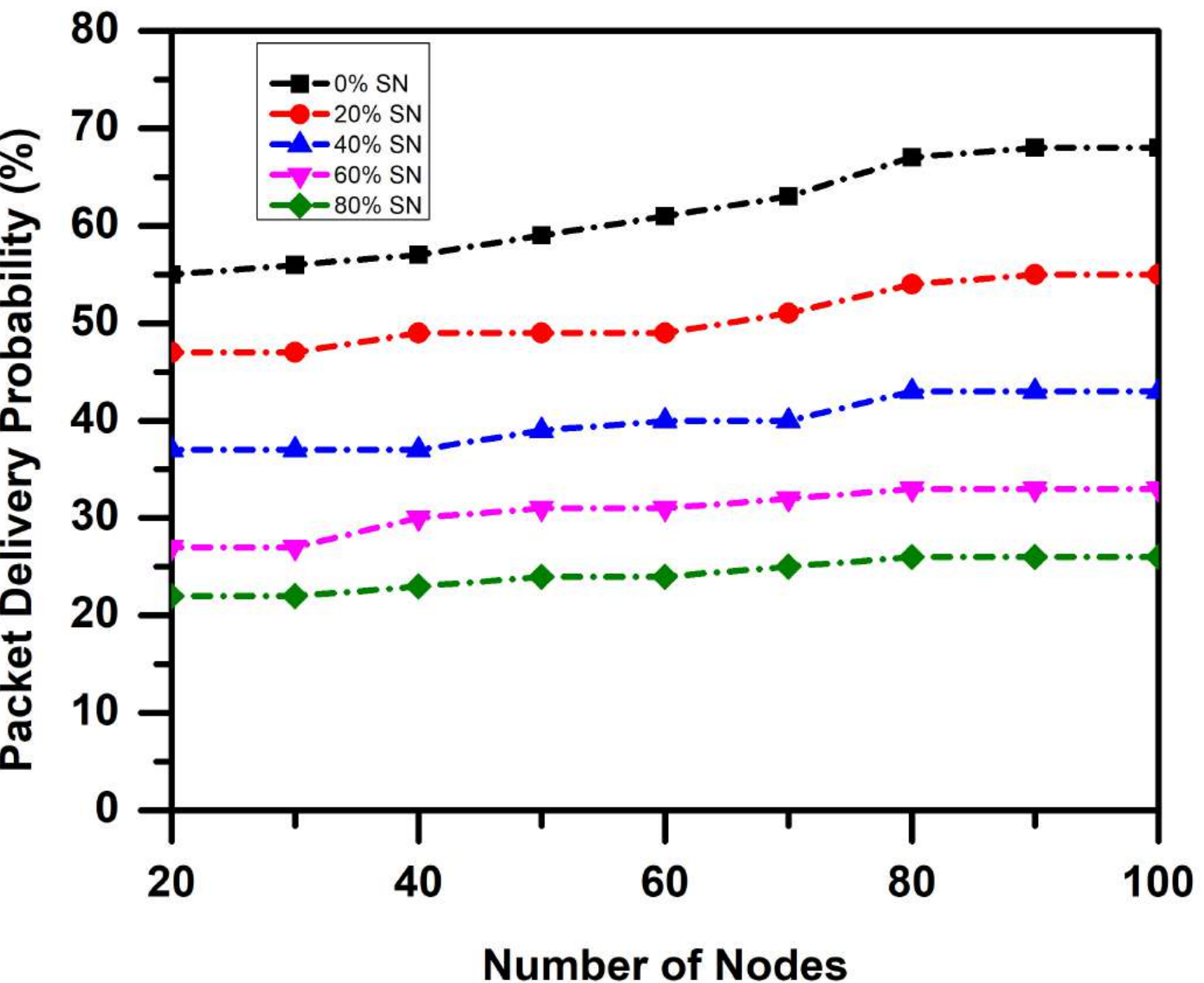}  
  \caption{Packet Delivery Probability}
  \label{fig:delivery1}
\end{subfigure}
\begin{subfigure}{.5\textwidth}
  \centering
  \includegraphics[scale =.43]{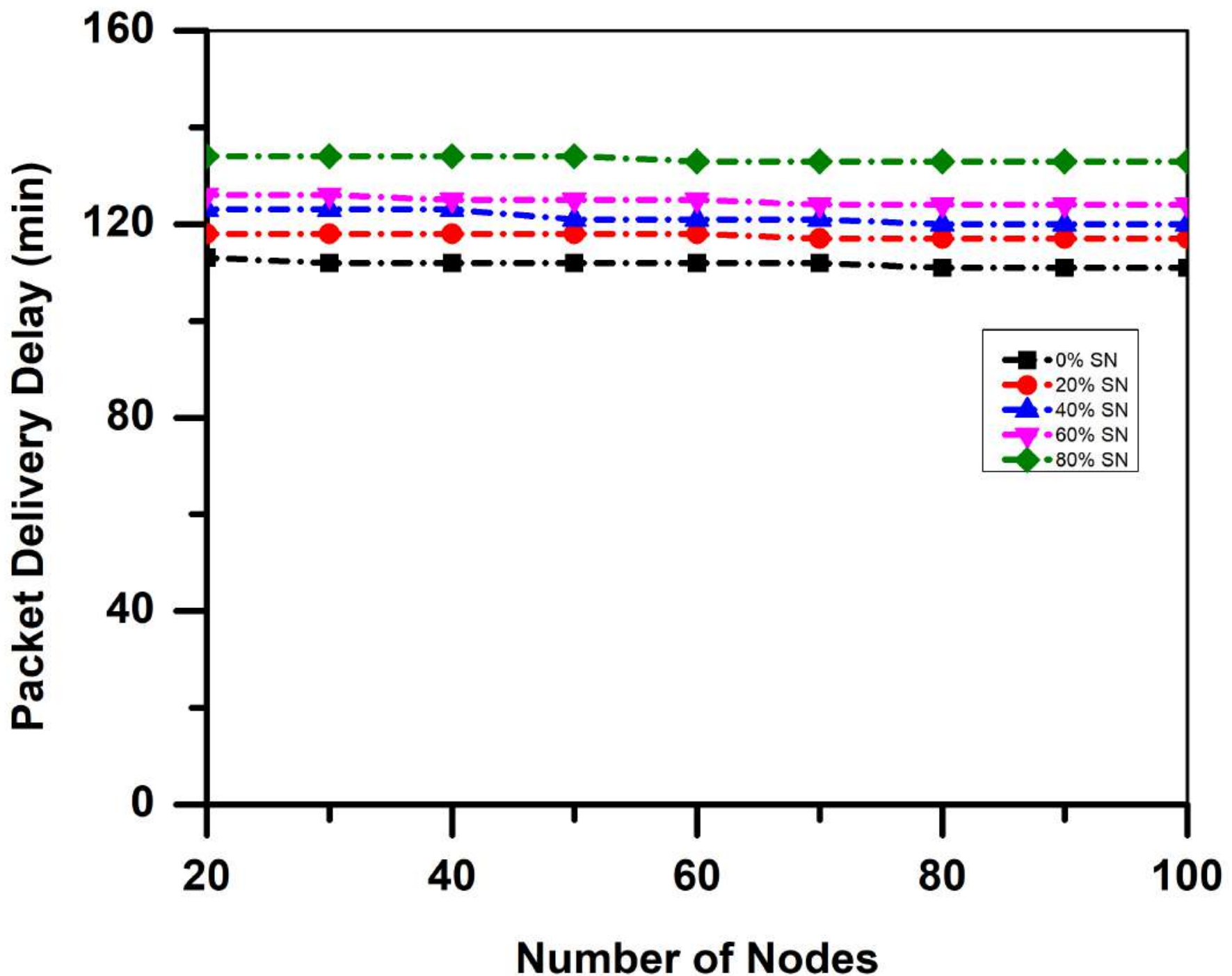}  
  \caption{Packet Delivery Delay}
  \label{fig:delay1}
\end{subfigure}
\begin{subfigure}{.5\textwidth}
  \centering
  \includegraphics[scale =.43]{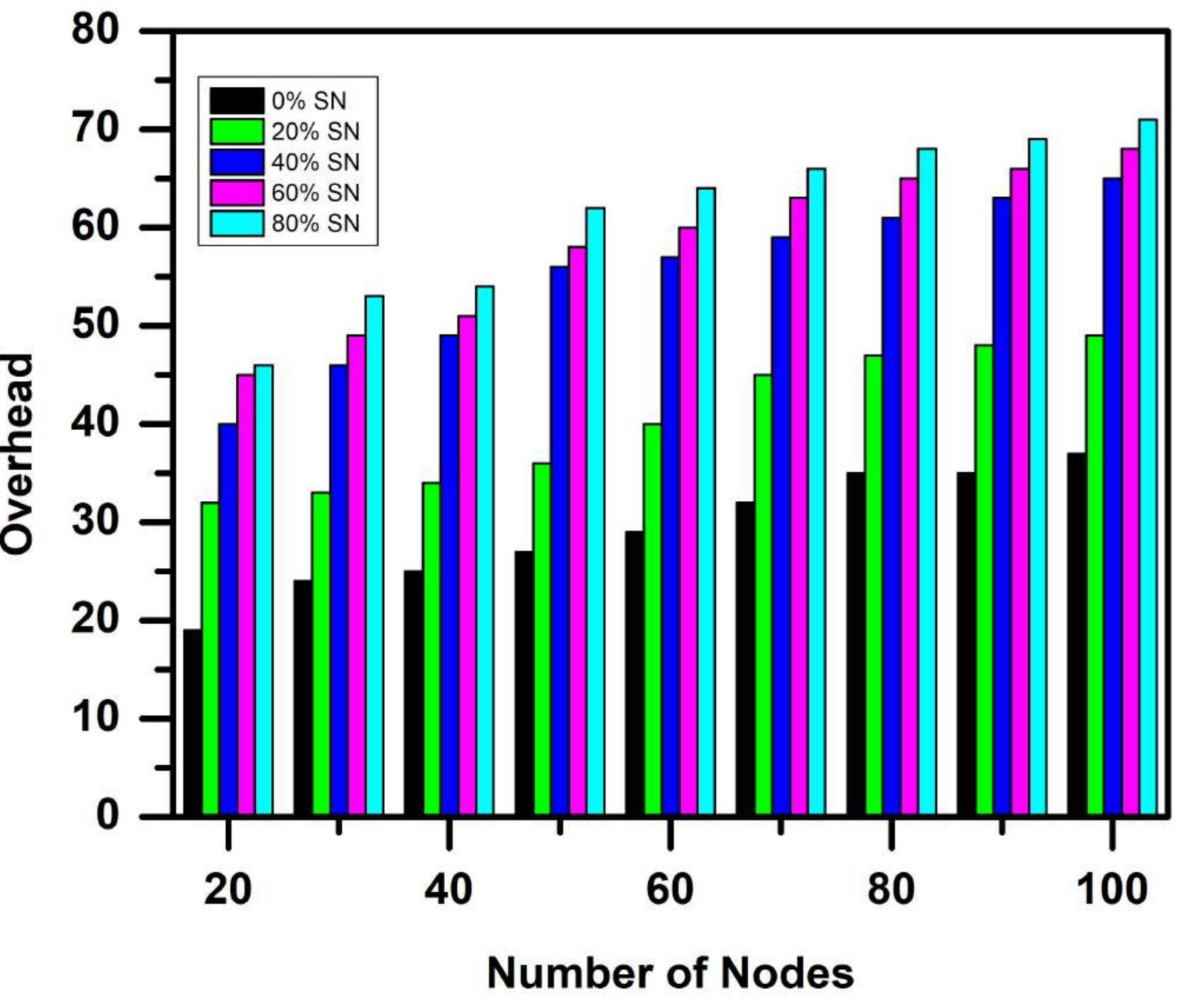}  
  \caption{Overhead Ratio}
  \label{fig:over1}
\end{subfigure}
\begin{subfigure}{.5\textwidth}
  \centering
  \includegraphics[scale =.43]{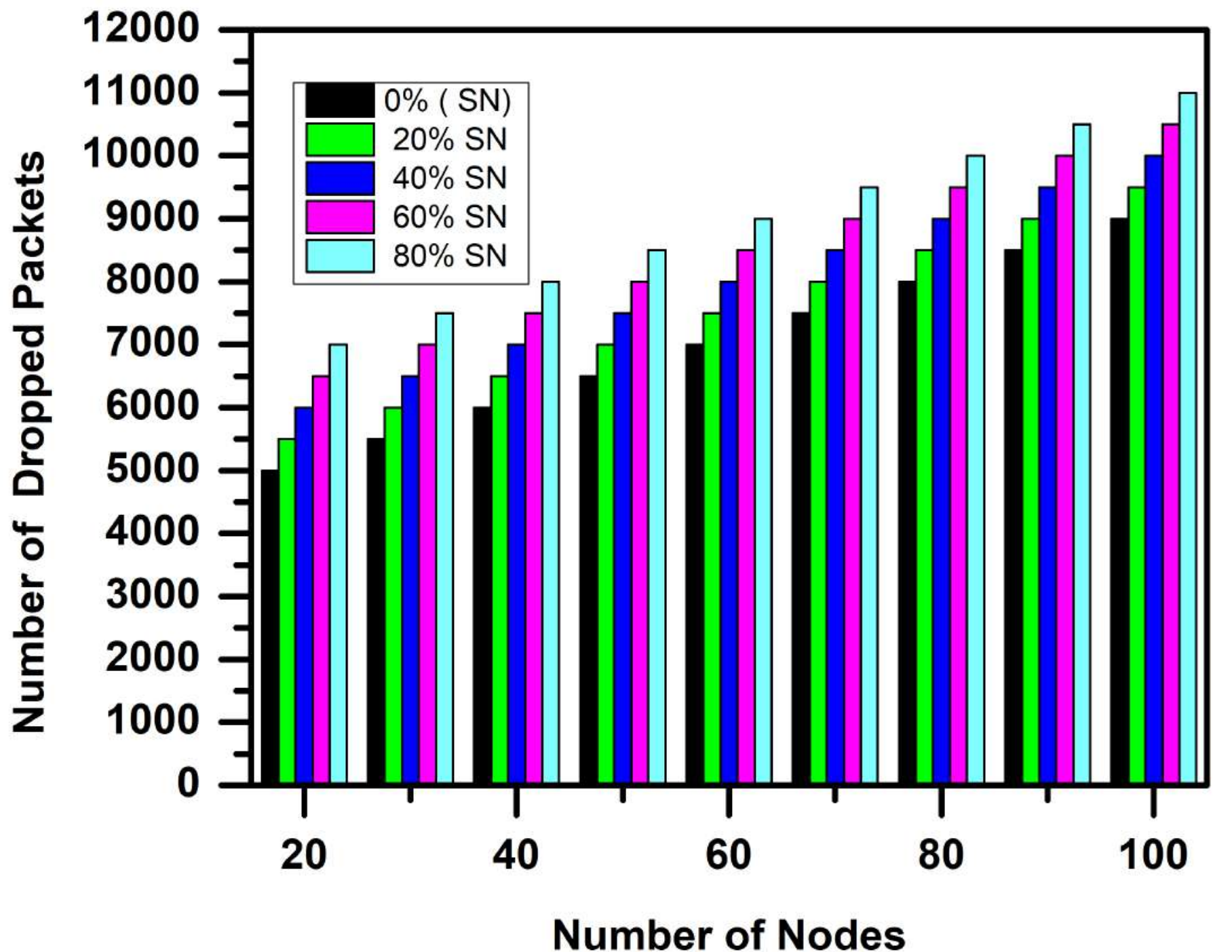}  
  \caption{Number of Dropped Packets}
  \label{fig:drop1}
\end{subfigure}
\caption{Influence of different number of selfish nodes on Performance Metrics}
\label{fig:fig1}
\end{figure}

The analysis begins with an evaluation of the influence of selfish nodes on all performance metrics such as packet delivery ratio, delivery delay, overhead ratio, and number of packet dropped as shown in figure~\ref{fig:fig1}. It is observed that the packets delivery ratio drastically decreases as the number of selfish nodes increases in the network as shown in~\ref{fig:delivery1}. This operation shows how important it is to detect and take steps against such nodes (e.g. punish or exclude them from the network). The selfish behavior of the nodes also affects the time packets take to reach its destination. It is due to the fact that the selfish nodes are not cooperating and it takes double effort from the cooperative nodes to deliver the packets to the destination. This process increases the amount of time a packet has to transfer between its source and the intended destination as shown in figure~\ref{fig:delay1}. 

By holding packets on node buffers for a prolonged period of time causes buffer congestion, resulting in a higher proportion of dropped packets as nodes must maintain their cooperative behavior so as to not deviate from the protocol. In addition to affecting other nodes, the existence of selfish nodes in the network will also have a massive impact on the routing process. The overhead is shown in figure~\ref{fig:over1} that describes the routing protocols bandwidth performance. It can be seen clearly when a number of selfish nodes increases it also increase the overhead that actually degrades the network performance. figure~\ref{fig:drop1} shows that the number of packets dropped as a result of the existence of selfish nodes.
It was observed throughout this section, as the percentage of selfish nodes increased it has drastically affected the performance of all the metrics. In next subsection, the performance of the proposed HBDS scheme is checked by incorporating the same percentage of selfish nodes in the network.

\subsubsection{Effect of Selfish Nodes on Packet Delivery Ratio and Average Delivery Delay} 
\begin{figure}[ht]
\begin{subfigure}{.5\textwidth}
  \centering
  \includegraphics[scale =.43]{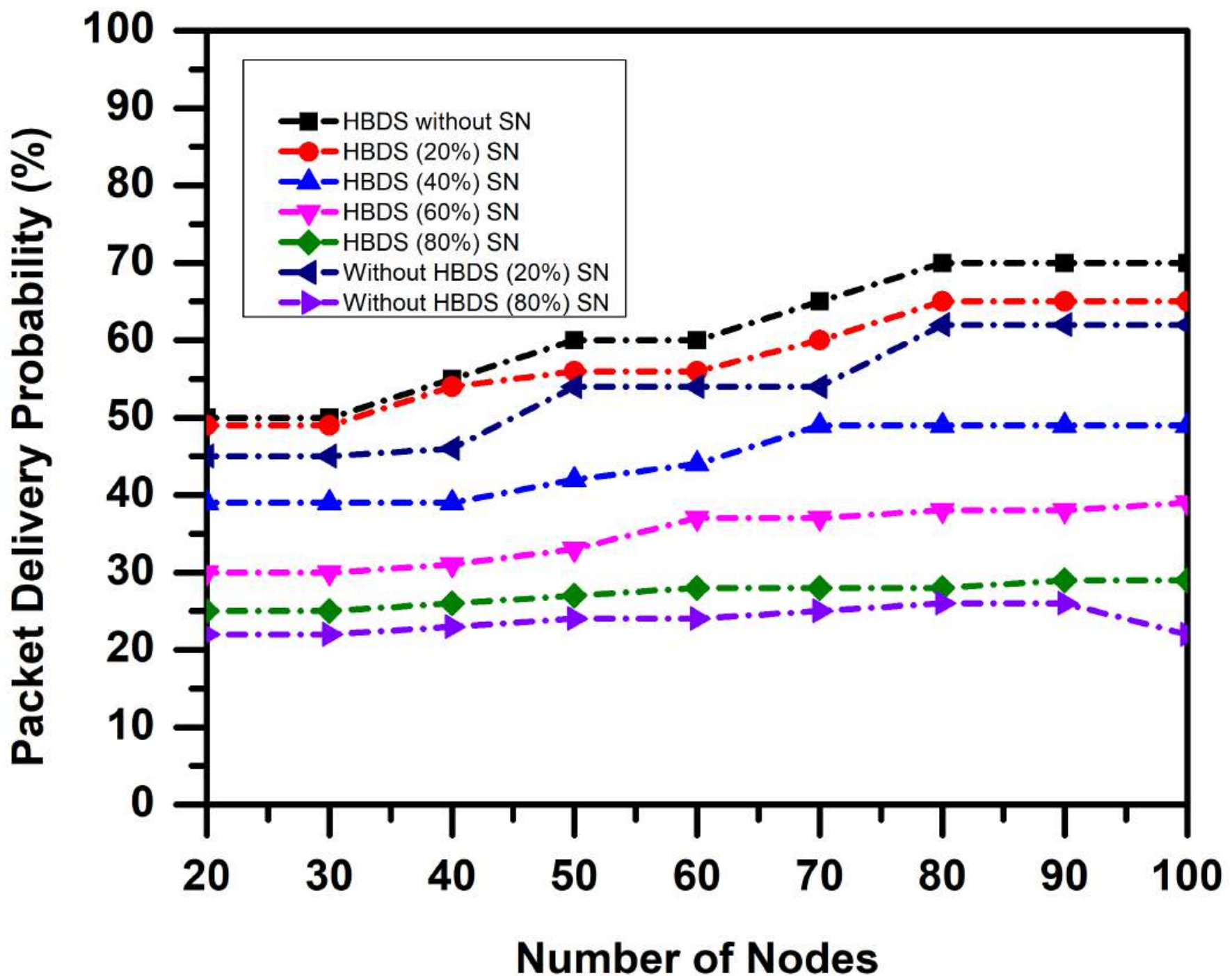}  
  \caption{Packet Delivery Probability}
  \label{fig:delivery2}
\end{subfigure}
\begin{subfigure}{.5\textwidth}
  \centering
  \includegraphics[scale =.43]{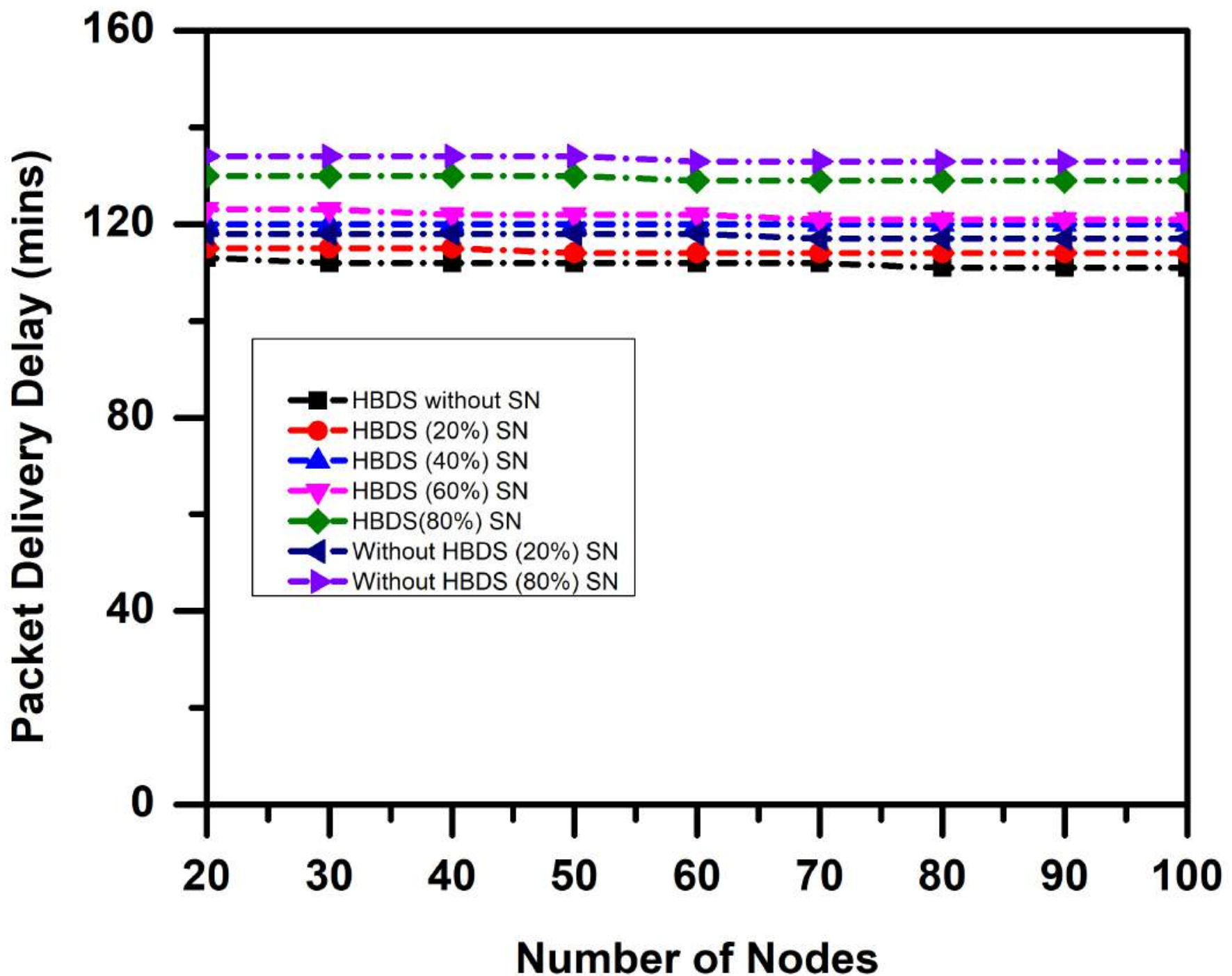}  
  \caption{Number of Dropped Packets}
  \label{fig:delay2}
\end{subfigure}
\caption{Packet Delivery and Delay when 20\% to 80\% nodes are selfish nodes}
\label{fig:fig2}
\end{figure}

To evaluate the performance of the proposed scheme, the result of the HBDS scheme was compared with a scheme where there is no selfishness omitting procedure was used as shown in figure~\ref{fig:fig2}. This study begins with an assessment of the percentage of packets delivered. As shown in figure~\ref{fig:delivery2}, as the number of selfish nodes increases, the probability of packet delivery continues to decrease. 

However, the HBDS succeeds in reducing the negative effect of selfish nodes. Comparing HBDS packet delivery outcomes with a strategy in which no action is taken toward selfish nodes. By incorporating 20\% selfish nodes in the network, the HBDS increases the packet delivery outcome by approximately 4\%, 4\%, 8\%, 2\%, 2\%, 6\%, 3\%, 3\% and 3\% (for a number of nodes 20, 30, 40, 50, 60, 70, 80, 90, and 100 respectively). In addition, by incorporating 80\% selfish nodes, HBDS also enables to reduce the effects of selfish nodes by increasing the probability of packet delivery. The packet delivery probability is increased approximately by 3\%, 3\%, 3\%, 3\%, 4\%, 3\%, 2\%, 3\% and 7\% (for number of nodes 20, 30, 40, 50, 60, 70, 80, 90, and 100 respectively. This is because, HBDS rewards nodes in the form incentive (reputation) for their cooperative behavior, making cooperative nodes even more resource-sharing.

It can also be seen in figure~\ref{fig:delay2} that HBDS also produces a good performance in terms of packet delivery delay when compared to a scheme in which there is no strategy used for omitting selfishness in a community. It ensures that HBDS is capable of delivering packets faster. When 20\% selfish nodes are incorporated in the network, the HBDS packets delivery approximately 5, 5, 5, 5, 6, 6, 6, 6, and 6 sooner (for a number of nodes 20, 30, 40, 50, 60, 70, 80, 90 and 100 respectively). Also, when 80\% selfish nodes are incorporated in the network, the HBDS packets delivery approximately 16, 16, 16, 16, 15, 16, 16, 16, and 16 sooner (for number of nodes 20, 30, 40, 50, 60, 70, 80, 90 and 100 respectively). The explanation for the observed results for both parameters is due to the HBDS approach of considering a reputation score of nodes to determine the number of resources they can share. The node reputation score will be increased by allowing only interaction between cooperative nodes, that is resulting in higher resource sharing.

\subsubsection{Effect of Selfish Nodes on Number of Packets Drops and Overhead Ratio} 

\begin{figure}[ht]
\begin{subfigure}{.5\textwidth}
  \centering
  \includegraphics[scale =.43]{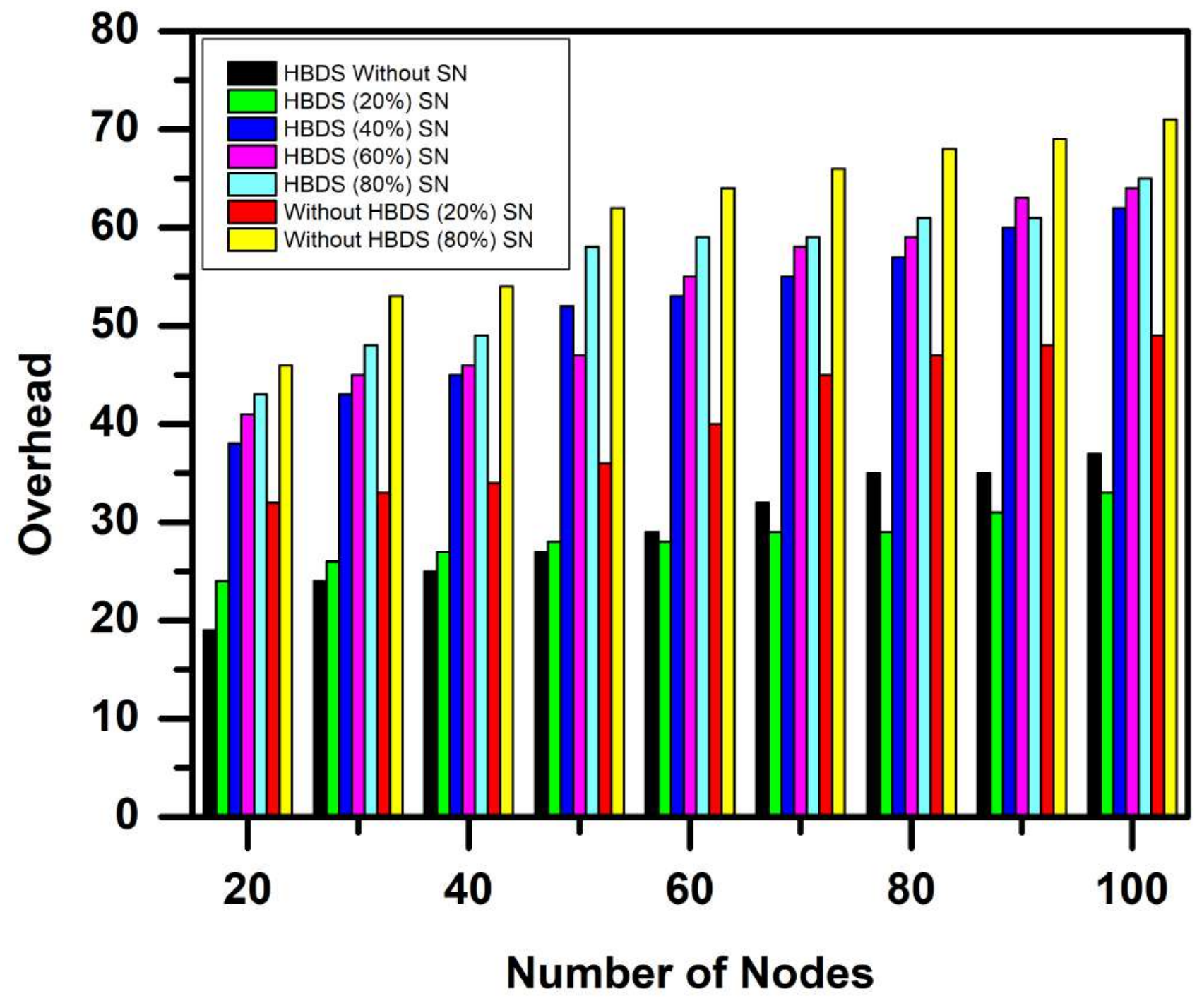}  
  \caption{Overhead Ratio}
  \label{fig:over2}
\end{subfigure}
\begin{subfigure}{.5\textwidth}
  \centering
  \includegraphics[scale =.43]{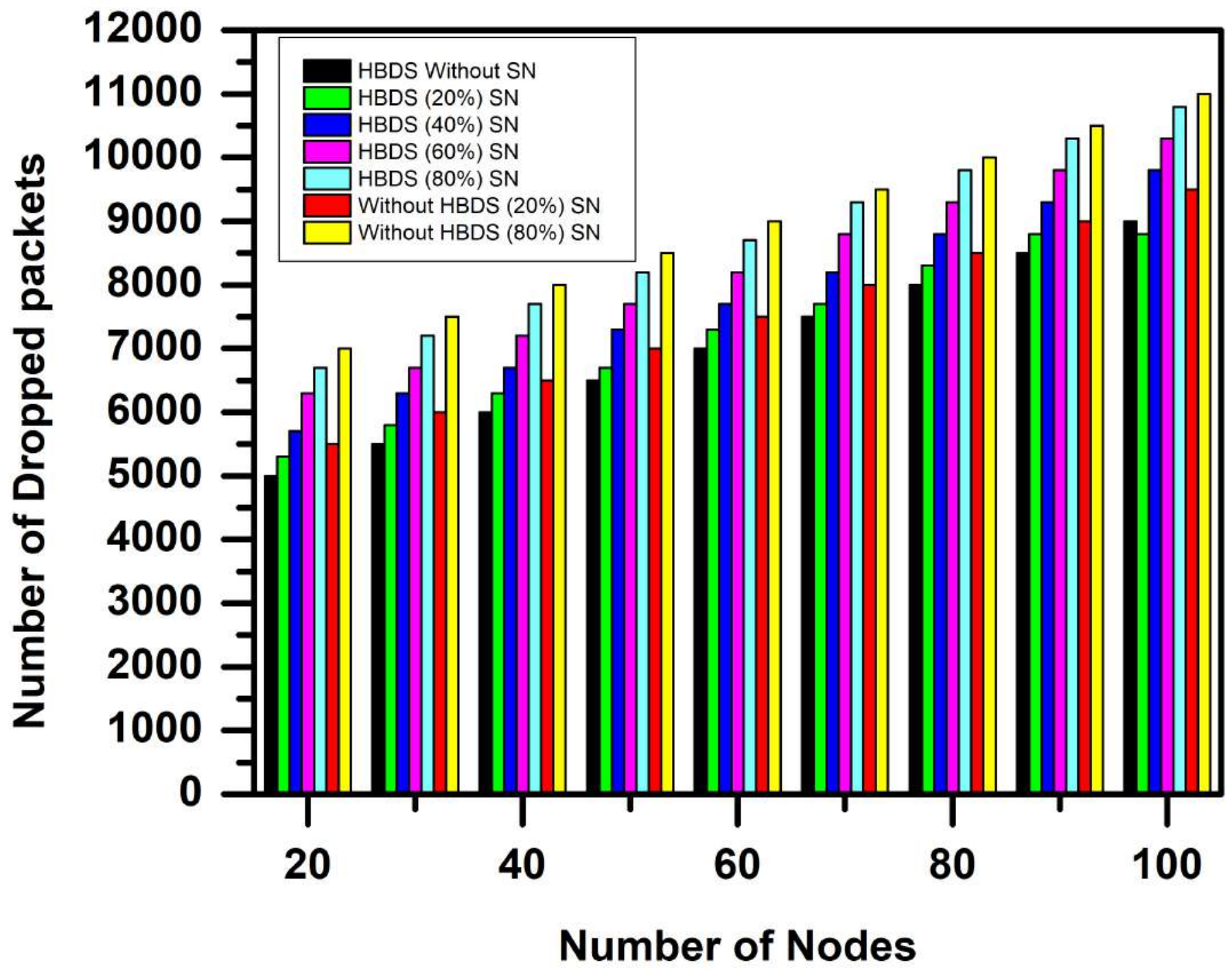}  
  \caption{Number of Dropped Packets}
  \label{fig:drop2}
\end{subfigure}
\caption{Overhead Ratio and Number of Dropped Packets when 20\% to 80\% nodes are selfish nodes}
\label{fig:fig3}
\end{figure}

The performance of the proposed scheme HBDS is checked in terms of wasted resources in this section. The evaluation begins with an assessment of the overhead ratio and number of packets dropped for this reason as shown in figure~\ref{fig:fig3}. It can be seen in figure~\ref{fig:over2}, the selfish nodes do not have a huge impact on the overhead ratio in the proposed scheme HBDS. 

This is a big achievement when the result of the proposed scheme HBDS is compared with the result of an approach where no action has been taken toward selfish nodes. For 20\% selfish nodes the overhead of HBDS is decreased approximately by 8, 7, 7, 8, 12, 16, 18, 17, and 16 packets (for a number of nodes 20, 30, 40, 50, 60, 70, 80, 90 and 100 respectively). Similarly for For 80\% selfish nodes, the overhead of the proposed scheme HBDS is reduced approximately by 3, 5, 5, 4, 5, 7, 7, 8, and 6 packets (for a number of nodes 20 , 30, 40, 50, 60, 70, 80, 90 and 100 respectively). The proposed scheme also significantly reduced the number of packets drops for different percentage of selfish nodes as shown in figure~\ref{fig:drop2}. 

For 20\% selfish nodes, the HBDS scheme dropping packets approximately 200, 200, 200, 300, 200, 300, 200, 200, and 700 (for a number of nodes 20, 30, 40, 50, 60, 70, 80, 90 and 100 respectively) less than the approach where there is no proper selfish nodes detection procedure used. In addition, for 80\% selfish nodes, the proposed scheme discard less 200, 200, 200, 300, 200, 300, 200, 200, and 700 packets (for a number of nodes 20, 30, 40, 50, 60, 70, 80, 90 and 100 respectively). This happens because HBDS helps to reduce the network resource due to the proper monitoring of selfish nodes in the community and avoiding contacts with such nodes. Thus, avoiding selfish node contact often means that a higher number of copies of packets pass through the network, improving their chances to reach the final destination.

\subsubsection{Comparison of HBDS, SSAR, and SimBet for Different Percentage of Selfish Nodes}

\begin{figure}[ht]
\begin{subfigure}{.5\textwidth}
  \centering
  \includegraphics[scale =.43]{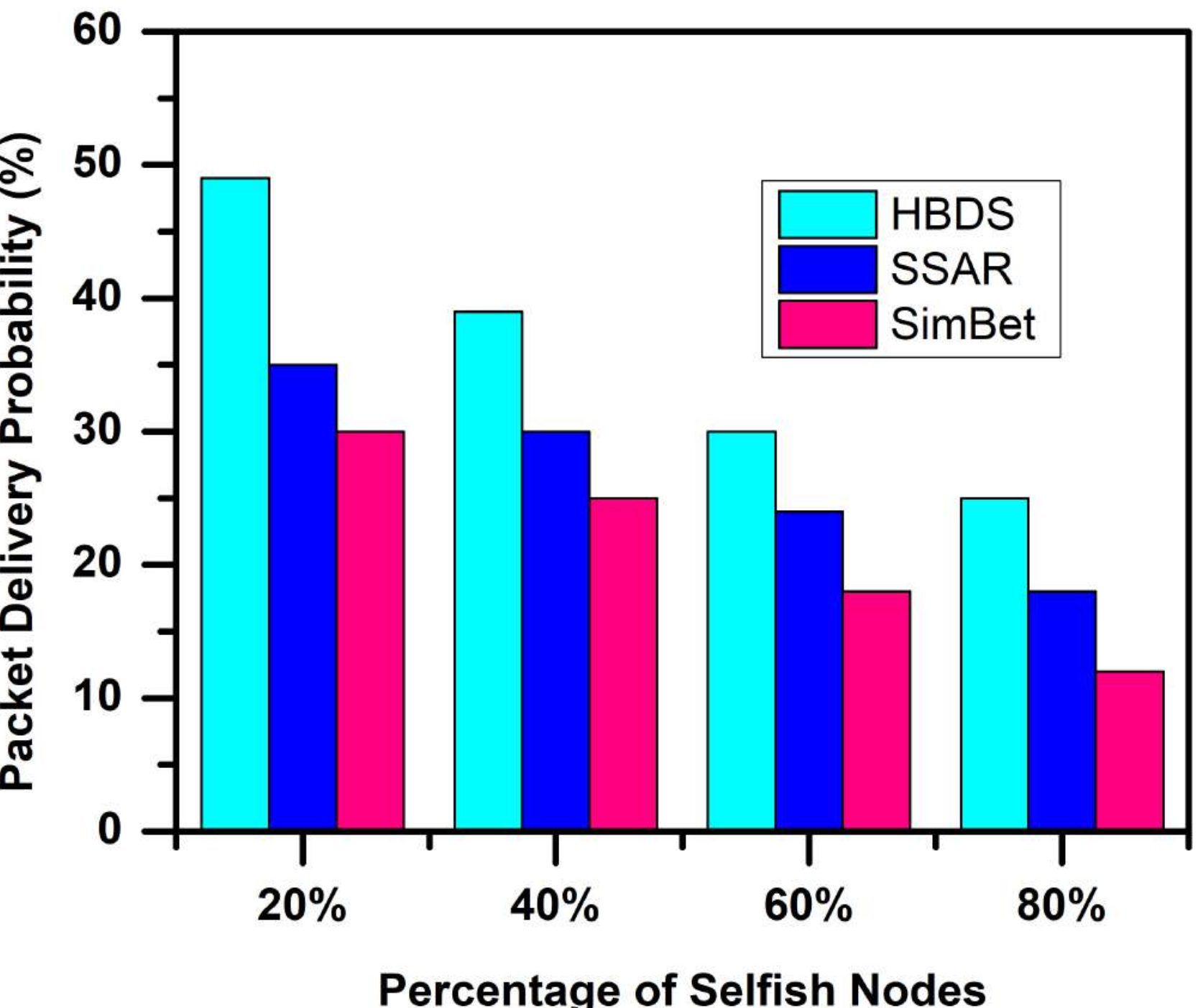}  
  \caption{Packet Delivery Probability}
  \label{fig:deliverycomp}
\end{subfigure}
\begin{subfigure}{.5\textwidth}
  \centering
  \includegraphics[scale =.43]{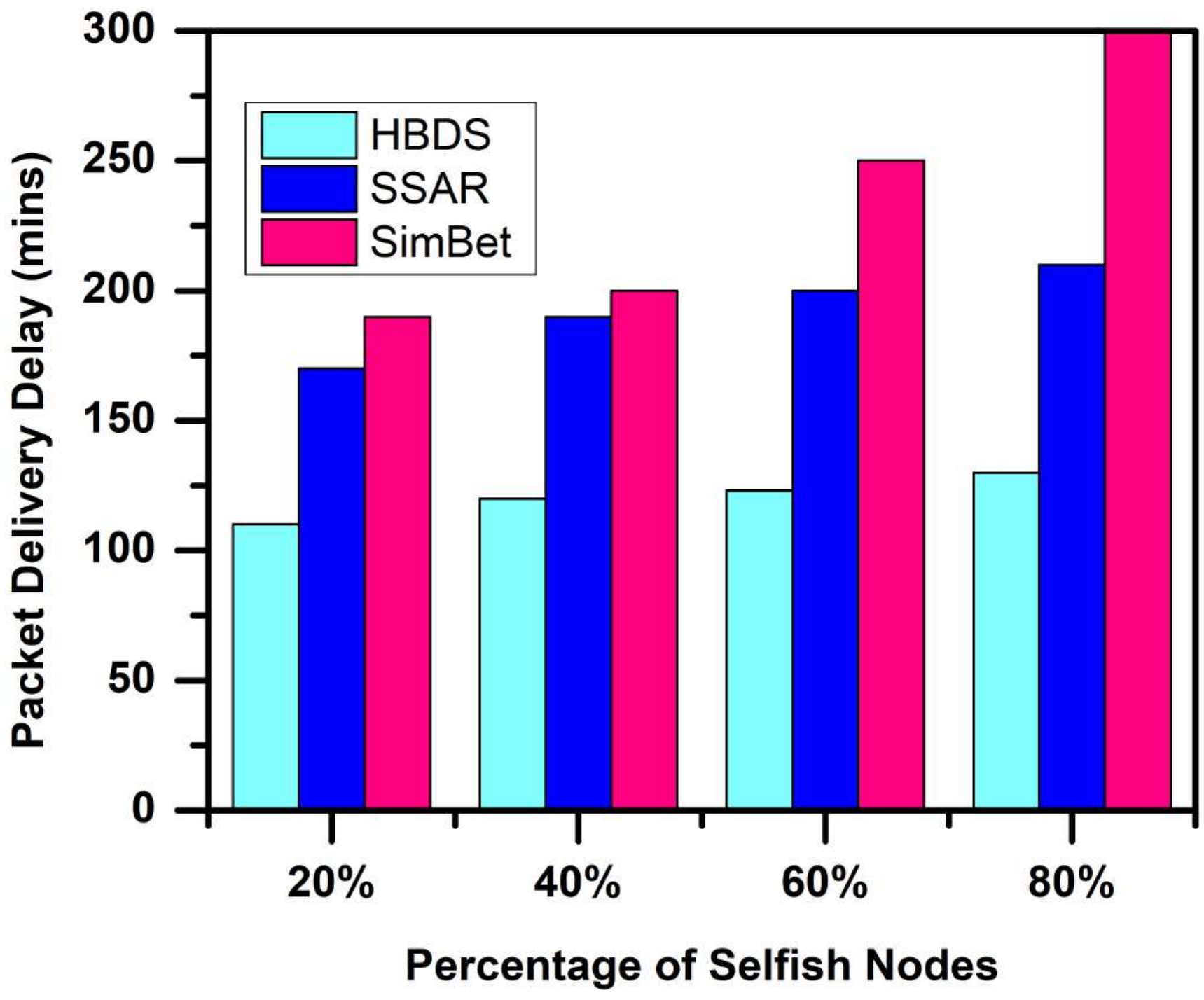}  
  \caption{Packet Delivery Delay}
  \label{fig:delaycomp}
\end{subfigure}
\begin{subfigure}{.5\textwidth}
  \centering
  \includegraphics[scale =.43]{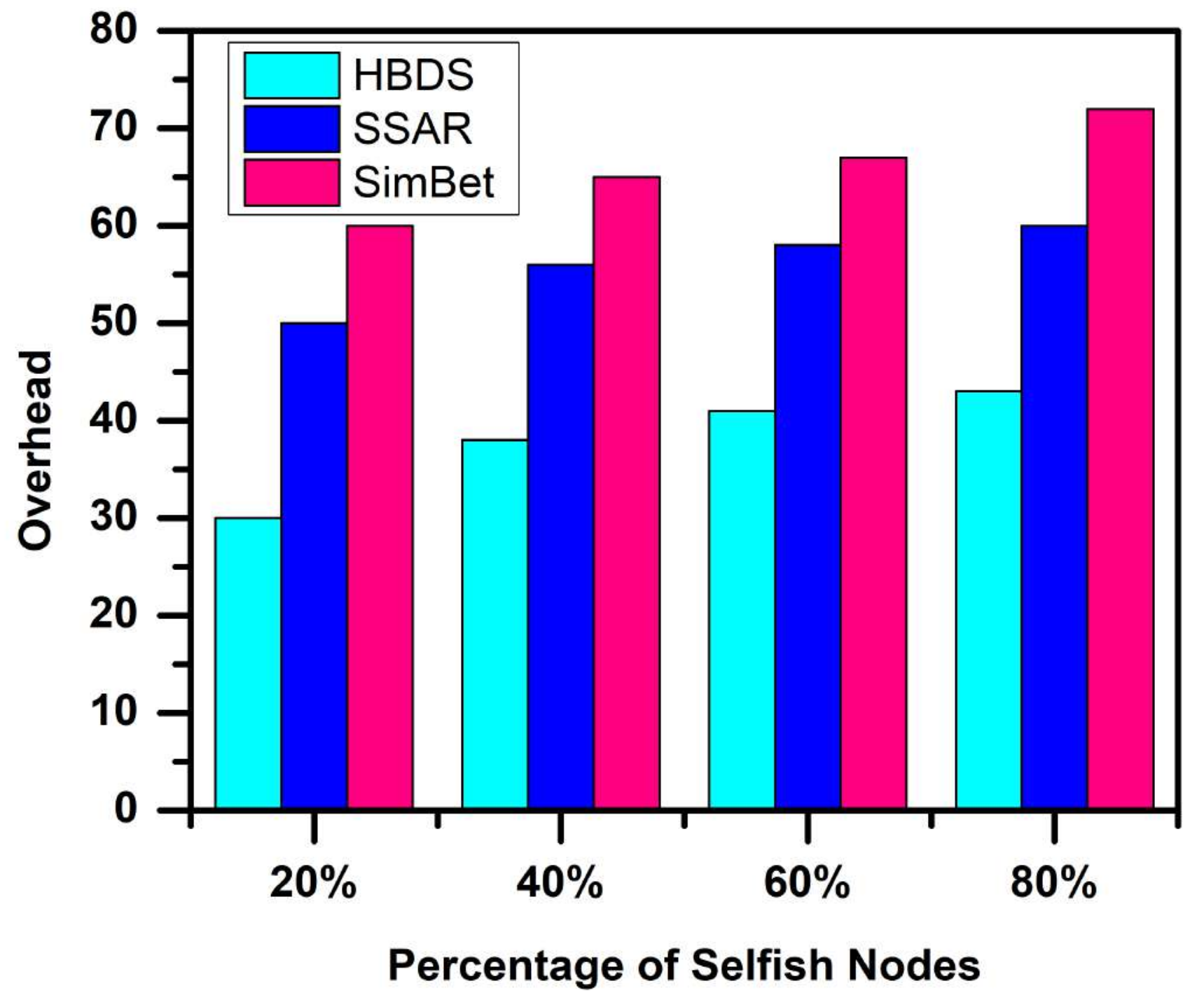}  
  \caption{Overhead Ratio}
  \label{fig:overcomp}
\end{subfigure}
\begin{subfigure}{.5\textwidth}
  \centering
  \includegraphics[scale =.43]{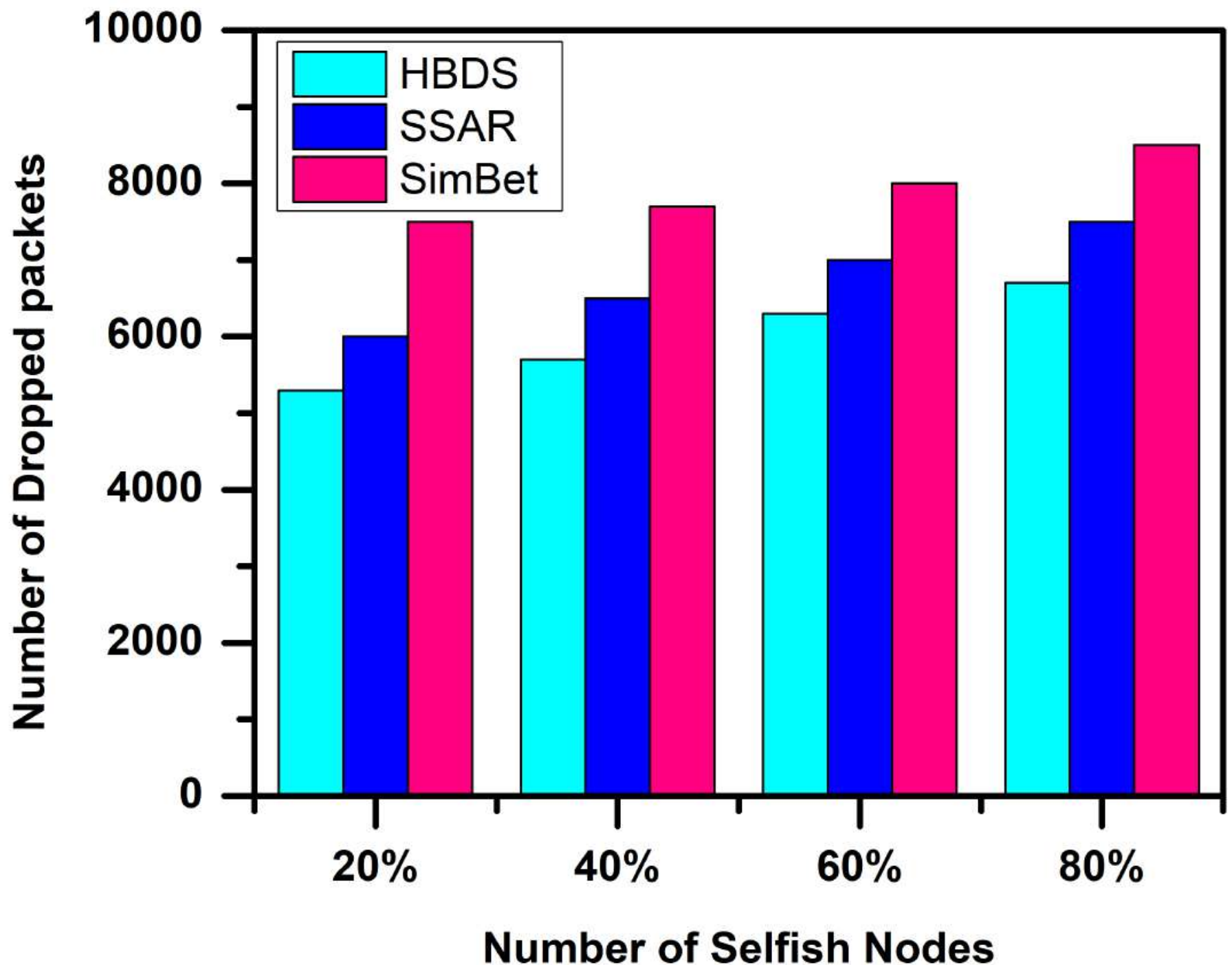}
  \caption{Number of Dropped Packets}
  \label{fig:dropcomp}
\end{subfigure}
\caption{Performance Comparison of HBDS, SSAR and SimBet for Selfish Nodes Percentage of 20\% to 80\% for all Metrics}
\label{fig:fig4}
\end{figure}

These three techniques are evaluated for the network characteristics in case of 20\%, 40\%, 60\%, and 80\% nodes are selfish. The HBDS scheme outperforms the two schemes SSAR and SimBet in terms of Packet delivery probability, packet delivery delay, overhead ratio and a number of dropped packets as shown in ~\ref{fig:fig4}. In figure~\ref{fig:deliverycomp}, the packet delivery probability of the proposed HBDS scheme is higher. 

For 20\% selfish nodes, the packet delivery probability of HBDS, SSAR, and SimBet is 49\%, 35\% and 30\% respectively that is 14\% and 19\% higher than SSAR and SimBet. Similarly for 80\% selfish nodes, the packet delivery probability of HBDS is 25\% that is 7\% and 13\% higher than SSAR and SimBet respectively. The key reason behind this is that the HBDS scheme encourages selfish nodes to take part in the routing process and thus avoid the selfish behavior of nodes. The other three techniques did not present in-depth the effect of selfishness on the network. In figure~\ref{fig:delaycomp}, the packet delivery delay, overhead ratio and a number of packets dropped of the Proposed HBDS scheme are lower. For 20\% selfish nodes, the packet delivery delay of HBDS is 110 minutes that is 20\% and 27\% lower than SSAR and SimBet respectively. By incorporating 80\% selfish nodes, the packet delivery delay of the HBDS is 130 minutes that is 27\% and 57\% lower than SSAR and SimBet respectively. In figure~\ref{fig:overcomp}, the overhead of HBDS scheme is 25\% and 37.5\% lower than SSAR and SimBet respectively when there are 20\% selfish nodes in the network.
 The comparison of HDBS, SSAR, and SimBet for packet delivery ratio for different percentage of selfish nodes is shown in Table~\ref{table3}. For 20\% selfish nodes, the packet delivery probability of HBDS, SSAR, and SimBet are 49, 35, and 30 respectively. Thus, it can be seen that the packet delivery ratio of HBDS is 23\% and 31\% higher than SSAR and SimBet respectively. In addition, for 80\% selfish nodes, the packet delivery probability of HBDS, SSAR, and SimBet are 25, 18, and 12 that is 11.6\% and 21.6\% higher than SSAR and SimBet respectively.
 
 The comparison of HDBS, SSAR, and SimBet for delivery delay for different percentage of selfish nodes is shown in Table~\ref{table4}. For 20\% selfish nodes, the delivery delay of HBDS, SSAR, and SimBet are 110, 170, and 190 respectively. Thus, it can be seen that the delivery delay of HBDS is 20\% and 27\% lower than SSAR and SimBet respectively. In addition, for 80\% selfish nodes, the delivery delay of HBDS, SSAR, and SimBet are 130, 210, and 300 that is 56\% and 26\% lower than SSAR and SimBet respectively.
\begin{table}
\center
\caption{Comparison of HBDS, SSAR, and SimBet in case of 20\%, 40\%, 60\%, and 80\% Selfish nodes for Packet Delivery Probability}
\label{table3}
\begin{tabular}{llll}
\toprule
 No. of Selfish Nodes & HBDS&SSAR &SimBet\\
\midrule
20\%&49&35&30\\
40\%&39&30&25\\
60\%&30&24&18\\
80\%&25&18&12\\
\bottomrule
 \end{tabular}
 \end{table} 

\begin{table}
\center
\caption{Comparison of HBDS, SSAR, and SimBet in case of 20\%, 40\%, 60\%, and 80\% Selfish nodes for Delivery Delay}
\label{table4}
\begin{tabular}{llll}
\toprule
 No. of Selfish Nodes & HBDS&SSAR &SimBet\\
\midrule
20\%&110&170&190\\
40\%&120&190&200\\
60\%&123&200&250\\
80\%&130&210&300\\
\bottomrule
 \end{tabular}
 \end{table} 

In addition, for 80\% selfish nodes, the overhead of proposed scheme is 22\% and 37\% lower than SSAR and SimBet respectively. In figure~\ref{fig:dropcomp}, for 20\% selfish nodes, the number of packet drop in HBDS scheme is 7\% and 22\% less than SSAR and SimBet respectively. The comparison of HDBS, SSAR, and SimBet for overhead ratio for different percentage of selfish nodes is shown in Table~\ref{table5}. For 20\% selfish nodes, the overhead ratio of HBDS, SSAR, and SimBet are 30, 50, and 60 respectively. Thus, it can be seen that the overhead ratio of HBDS is 25\% and 38\% lower than SSAR and SimBet respectively. In addition, for 80\% selfish nodes, the overhead of HBDS, SSAR, and SimBet are 43, 60, and 72 that is 22\% and 37\% lower than SSAR and SimBet respectively. The comparison of HDBS, SSAR, and SimBet for packet delivery ratio for different percentage of selfish nodes is shown in Table~\ref{table6}. For 20\% selfish nodes, the rate of packet drop of HBDS, SSAR, and SimBet are 5300, 6000, and 7500 respectively. Thus, it can be seen that the overhead ratio of HBDS is 7\% and 22\% lower than SSAR and SimBet respectively. In addition, for 80\% selfish nodes, the rate of packet drop of HBDS, SSAR, and SimBet are 6700, 7500, and 8500 respectively, that is 8\% and 18\% lower than SSAR and SimBet respectively. It is because the nodes avoid sending messages to their neighboring nodes in other methods, if they do not have a strong social interaction with their neighbors. The nodes are considered to be cooperative residing in the same community in SimBet and SAAR methods. So in this protocol, the level of selfishness is low. While the nodes will modify their selfishness accordingly in our HBDS scheme if they have sufficient incentive for data transfer. This improves network performance by changing the selfishness of the node dynamically. Therefore, it concludes that HBDS scheme has a better capacity to handle node selfishness and to encourage nodes to collaborate in a better network performance.   
\begin{table}
\center
\caption{Comparison of HBDS, SSAR, and SimBet in case of 20\%, 40\%, 60\%, and 80\% Selfish nodes for Overhead Ratio}
\label{table5}
\begin{tabular}{llll}
\toprule
 No. of Selfish Nodes & HBDS&SSAR &SimBet\\
\midrule
20\%&30&50&60\\
40\%&38&56&65\\
60\%&41&58&67\\
80\%&43&60&72\\
\bottomrule
 \end{tabular}
 \end{table} 
 
 \begin{table}
 \center
\caption{Comparison of HBDS, SSAR, and SimBet in case of 20\%, 40\%, 60\%, and 80\% Selfish nodes for Number of Packets Drop}
\label{table6}
\begin{tabular}{llll}
\toprule
 No. of Selfish Nodes & HBDS&SSAR &SimBet\\
\midrule
20\%&5300&6000&7500\\
40\%&5700&6500&7700\\
60\%&6300&7000&8000\\
80\%&6700&7500&8500\\
\bottomrule
 \end{tabular}
 \end{table}
\section{Conclusion and Future Work}
\label{sec:con}
\noindent
In this article, the HBDS scheme is proposed for IoT using VDTN to stimulate selfish nodes in a community for cooperation or forwarding of messages. The HBDS scheme monitors selfish nodes and continuously watches their behavior. The reward is offered as a reputation that makes the network node cooperative. Nodes are also penalized for showing consistently selfish behavior. Following the democratic process, the watchdog nodes evaluate the performance of their neighbor nodes. Two protocols namely SSAR and SimBet were systematically analyzed to test the performance metrics such as packet delivery probability, packet delivery delay, overhead ratio and number of dropped packets. The results show that the HBDS scheme can effectively manage a wide range of selfish nodes by enabling more selfish nodes to work together in a community to boost network performance.

The proposed scheme HBDS can be used in the future for cooperative sharing of the information among the connected nodes in the network. It can be used for efficient tracking of the selfish nodes to detect the network anomalies like errors in packets delivery, damaged packets delivered, energy and power constraints.

\bibliography{MyBibtex.bib}%

\begin{thebibliography}{10}
\providecommand \doibase [0]{http://dx.doi.org/}%

\bibitem{p00}
Benhamida FZ, Bouabdellah A, Challal Y. Using delay tolerant network for the
  internet of things: Opportunities and challenges. In:  {\it 2017 8th
  International Conference on Information and Communication Systems
  (ICICS)}IEEE. ; 2017\string: 252--257.

\bibitem{p01}
Khabbaz MJ, Assi CM, Fawaz WF. Disruption-tolerant networking: A comprehensive
  survey on recent developments and persisting challenges. {\it IEEE
  Communications Surveys \& Tutorials} 2011\string; 14(2)\string: 607--640.

\bibitem{p02}
Ahmed SH, Kang H, Kim D. Vehicular delay tolerant network (VDTN): Routing
  perspectives. In:  {\it 2015 12th Annual IEEE Consumer Communications and
  Networking Conference (CCNC)}IEEE. ; 2015\string: 898--903.

\bibitem{p03}
Kang H, Ahmed SH, Kim D, Chung YS. Routing protocols for vehicular delay
  tolerant networks: a survey. {\it International Journal of Distributed Sensor
  Networks} 2015\string; 11(3)\string: 325027.

\bibitem{p04}
Benamar M, Ahnana S, Saiyari FZ, Benamar N, El~Ouadghiri MD, Bonnin JM. Study
  of VDTN Routing Protocols Performances in Sparse and Dense Traffic in the
  Presence of Relay Nodes.. {\it J. Mobile Multimedia} 2014\string;
  10(1\&2)\string: 78--93.

\bibitem{p05}
Cherif AH, Boussetta K, Diaz G, Fedoua D. Improving the performances of
  geographic VDTN routing protocols. In:  {\it 2017 16th Annual Mediterranean
  Ad Hoc Networking Workshop (Med-Hoc-Net)}IEEE. ; 2017\string: 1--5.

\bibitem{p06}
Dias JA, Rodrigues JJ, Xia F, Mavromoustakis CX. A cooperative watchdog system
  to detect misbehavior nodes in vehicular delay-tolerant networks. {\it IEEE
  Transactions on Industrial Electronics} 2015\string; 62(12)\string:
  7929--7937.

\bibitem{p07}
Raw RS, Kadam A, others . Performance Analysis of DTN Routing Protocol for
  Vehicular Sensor Networks. In:  {\it Next-Generation Networks}Springer.  2018
  (pp. 229--238).

\bibitem{p07a}
Khiadani NH, Safavi~Hemami SM, Hendessi F. Introducing a handshake for data
  dissemination using rateless codes in vehicular ad hoc networks. {\it
  Transactions on Emerging Telecommunications Technologies} 2017\string;
  28(2)\string: e2944.

\bibitem{p08}
Dias JA, Rodrigues JJ, Kumar N, Mavromoustakis CX. A hybrid system to stimulate
  selfish nodes to cooperate in vehicular delay-tolerant networks. In:  {\it
  2015 IEEE International Conference on Communications (ICC)}IEEE. ;
  2015\string: 5910--5915.

\bibitem{p09}
Dias JA, Rodrigues JJ, Kumar N, Saleem K. Cooperation strategies for vehicular
  delay-tolerant networks. {\it IEEE Communications Magazine} 2015\string;
  53(12)\string: 88--94.

\bibitem{p10}
Dias JA, Rodrigues JJ, Shu L, Ullah S. A reputation system to identify and
  isolate selfish nodes in vehicular delay-tolerant networks. In:  {\it 2013
  13th International Conference on ITS Telecommunications (ITST)}IEEE. ;
  2013\string: 133--138.

\bibitem{p11}
Rehman GU, Ghani A, Zubair M, Naqvi SHA, Singh D, Muhammad S. IPS: Incentive
  and Punishment Scheme for Omitting Selfishness in the Internet of Vehicles
  (Iov). {\it IEEE Access} 2019\string; 7\string: 109026--109037.

\bibitem{p12}
Mohammed N, Otrok H, Wang L, Debbabi M, Bhattacharya P. A mechanism
  design-based multi-leader election scheme for intrusion detection in manet.
  In:  {\it 2008 IEEE Wireless Communications and Networking Conference}IEEE. ;
  2008\string: 2816--2821.

\bibitem{p12a}
Bibri SE. The IoT for smart sustainable cities of the future: An analytical
  framework for sensor-based big data applications for environmental
  sustainability. {\it Sustainable Cities and Society} 2018\string; 38\string:
  230--253.

\bibitem{p12b}
Sun Y, Song H, Jara AJ, Bie R. Internet of things and big data analytics for
  smart and connected communities. {\it IEEE access} 2016\string; 4\string:
  766--773.

\bibitem{p12b1}
Zheng Q, Zheng K, Chatzimisios P, Long H, Liu F. A novel link allocation method
  for vehicle-to-vehicle-based relaying networks. {\it Transactions on Emerging
  Telecommunications Technologies} 2016\string; 27(1)\string: 64--73.

\bibitem{p12c}
Bennati S, Pournaras E. Privacy-enhancing aggregation of Internet of Things
  data via sensors grouping. {\it Sustainable cities and society} 2018\string;
  39\string: 387--400.

\bibitem{p12d}
Khatua PK, Ramachandaramurthy VK, Kasinathan P, Yong JY, Pasupuleti J,
  Rajagopalan A. Application and Assessment of Internet of Things toward the
  Sustainability of Energy Systems: Challenges and Issues. {\it Sustainable
  Cities and Society} 2019\string: 101957.

\bibitem{p12e}
Fabian P, Rachedi A, Gu{\'e}guen C. Programmable objective function for data
  transportation in the Internet of Vehicles. {\it Transactions on emerging
  telecommunications technologies} 2020\string: e3882.

\bibitem{p12f}
Xie K, Xie K, He S, Zhang D, Wen J, Lloret J. Busy tone-based channel access
  control for cooperative communication. {\it Transactions on Emerging
  Telecommunications Technologies} 2015\string; 26(10)\string: 1173--1188.

\bibitem{p13}
Dubey BB, Chauhan N, Chand N, Awasthi LK. Incentive based scheme for improving
  data availability in vehicular ad-hoc networks. {\it Wireless Networks}
  2017\string; 23(6)\string: 1669--1687.

\bibitem{p13a}
Velivasaki THN, Karkazis P, Zahariadis TV, Trakadas PT, Capsalis CN.
  Trust-aware and link-reliable routing metric composition for wireless sensor
  networks. {\it Transactions on Emerging Telecommunications Technologies}
  2014\string; 25(5)\string: 539--554.

\bibitem{p14}
Xu S, Li M, Chen Y, Shu L, Gu X. A cooperation scheme based on reputation for
  opportunistic networks. In:  {\it 2013 International Conference on Computing,
  Management and Telecommunications (ComManTel)}IEEE. ; 2013\string: 289--294.

\bibitem{p15}
Jiang Q, Men C, Yu H, Cheng X. A secure credit-based incentive scheme for
  opportunistic networks. In:  {\it 2015 7th International Conference on
  Intelligent Human-Machine Systems and Cybernetics}. 1. IEEE. ; 2015\string:
  87--91.

\bibitem{p16}
Ning Z, Liu L, Xia F, Jedari B, Lee I, Zhang W. CAIS: A copy adjustable
  incentive scheme in community-based socially aware networking. {\it IEEE
  Transactions on Vehicular Technology} 2016\string; 66(4)\string: 3406--3419.

\bibitem{p17}
Park Y, Sur C, Rhee KH. A secure incentive scheme for vehicular delay tolerant
  networks using cryptocurrency. {\it Security and Communication Networks}
  2018\string; 2018.

\bibitem{p18}
Jiang Q, Men C, Tian Z. A Credit-Based Congestion-Aware Incentive Scheme for
  DTNs. {\it Information} 2016\string; 7(4)\string: 71.

\bibitem{p19}
Jedari B, Liu L, Qiu T, Rahim A, Xia F. A game-theoretic incentive scheme for
  social-aware routing in selfish mobile social networks. {\it Future
  Generation Computer Systems} 2017\string; 70\string: 178--190.

\bibitem{p20}
Dias JA, Rodrigues JJ, Shu L, Ullah S. Performance evaluation of a cooperative
  reputation system for vehicular delay-tolerant networks. {\it EURASIP Journal
  on Wireless Communications and Networking} 2014\string; 2014(1)\string: 88.

\bibitem{p21}
Mantas N, Louta M, Karapistoli E, Karetsos GT, Kraounakis S, Obaidat MS.
  Towards an incentive-compatible, reputation-based framework for stimulating
  cooperation in opportunistic networks: a survey. {\it IET Networks}
  2017\string; 6(6)\string: 169--178.

\bibitem{p22}
Jedari B, Xia F, Chen H, Das SK, Tolba A, Zafer AM. A social-based watchdog
  system to detect selfish nodes in opportunistic mobile networks. {\it Future
  Generation Computer Systems} 2019\string; 92\string: 777--788.

\bibitem{p23}
Kou M, Zhao Y, Cai H, Fan X. Study of a Routing Algorithm of Internet of
  Vehicles Based on Selfishness. In:  {\it 2018 IEEE International Conference
  on Smart Internet of Things (SmartIoT)}IEEE. ; 2018\string: 34--39.

\bibitem{p24}
Sharma R, Gupta D. A Reputation-Based Mechanism to Detect Selfish Nodes in
  DTNs. In:  {\it International Conference on Communications and Cyber Physical
  Engineering 2018}Springer. ; 2018\string: 55--61.

\bibitem{p25}
Cai Y, Fan Y, Wen D. An incentive-compatible routing protocol for two-hop
  delay-tolerant networks. {\it IEEE Transactions on Vehicular Technology}
  2015\string; 65(1)\string: 266--277.

\bibitem{p26}
Liu L, Yang Q, Kong X, et al. Com-bis: a community-based barter incentive
  scheme in socially aware networking. {\it International Journal of
  Distributed Sensor Networks} 2015\string; 11(8)\string: 671012.

\bibitem{p27}
Zhou H, Wu J, Zhao H, Tang S, Chen C, Chen J. Incentive-driven and
  freshness-aware content dissemination in selfish opportunistic mobile
  networks. {\it IEEE Transactions on Parallel and Distributed Systems}
  2014\string; 26(9)\string: 2493--2505.

\bibitem{p28}
Butty{\'a}n L, D{\'o}ra L, F{\'e}legyh{\'a}zi M, Vajda I. Barter trade improves
  message delivery in opportunistic networks. {\it Ad Hoc Networks}
  2010\string; 8(1)\string: 1--14.

\bibitem{p29}
Li Q, Gao W, Zhu S, Cao G. A routing protocol for socially selfish delay
  tolerant networks. {\it Ad Hoc Networks} 2012\string; 10(8)\string:
  1619--1632.

\bibitem{p30}
Liu J, Bic L, Gong H, Zhan S. Data collection for mobile crowdsensing in the
  presence of selfishness. {\it EURASIP journal on wireless communications and
  networking} 2016\string; 2016(1)\string: 82.

\bibitem{p31}
Fawaz W. Effect of non-cooperative vehicles on path connectivity in vehicular
  networks: A theoretical analysis and UAV-based remedy. {\it Vehicular
  Communications} 2018\string; 11\string: 12--19.

\bibitem{p32}
Li X, Wang J. A Generous Cooperative Routing Protocol for Vehicle-to-Vehicle
  Networks.. {\it TIIS} 2016\string; 10(11)\string: 5322--5342.

\bibitem{p33}
Socievole A, Caputo A, De~Rango F, Fazio P. Routing in Mobile Opportunistic
  Social Networks with Selfish Nodes. {\it Wireless Communications and Mobile
  Computing} 2019\string; 2019.

\bibitem{p34}
Umar MM, Khan S, Ahmad R, Singh D. Game theoretic reward based adaptive data
  communication in wireless sensor networks. {\it IEEE Access} 2018\string;
  6\string: 28073--28084.

\bibitem{p35}
Vamsi PR, Kant K. Trust aware cooperative routing method for WANETs. {\it
  Security and Communication Networks} 2016\string; 9(18)\string: 6189--6201.

\bibitem{p36}
Kumar S, Dutta K. Trust Based Intrusion Detection Technique to Detect Selfish
  Nodes in Mobile Ad Hoc Networks. {\it Wireless Personal Communications}
  2018\string; 101(4)\string: 2029--2052.

\bibitem{p37}
Pal NN, Gaikwad K, Das D. Trust Calculation and Route Discovery for Delay
  Tolerant Networks. In:  {\it TENCON 2018-2018 IEEE Region 10 Conference}IEEE.
  ; 2018\string: 1533--1537.

\bibitem{p38}
Dhurandher SK, Kumar A, Obaidat MS. Cryptography-based misbehavior detection
  and trust control mechanism for opportunistic network systems. {\it IEEE
  Systems Journal} 2017\string; 12(4)\string: 3191--3202.

\bibitem{p39}
Truong NB, Um TW, Lee GM. A reputation and knowledge based trust service
  platform for trustworthy social internet of things. {\it Innovations in
  clouds, internet and networks (ICIN)} 2016.

\bibitem{p40}
Mas-Colell A, Whinston MD, Green JR, others . {\it Microeconomic theory}. 1.
\newblock Oxford university press New York .
\newblock 1995.

\bibitem{p41}
Mohammed N, Otrok H, Wang L, Debbabi M, Bhattacharya P. Mechanism design-based
  secure leader election model for intrusion detection in MANET. {\it IEEE
  transactions on dependable and secure computing} 2009\string; 8(1)\string:
  89--103.

\bibitem{p42}
Sulaiman A, Raja SK, Park SH. Improving scalability in vehicular communication
  using one-way hash chain method. {\it Ad Hoc Networks} 2013\string;
  11(8)\string: 2526--2540.

\bibitem{p43}
Soares VN, Farahmand F, Rodrigues JJ. VDTNsim: a simulation tool for vehicular
  delay-tolerant networks. In:  {\it 2010 15th IEEE International Workshop on
  Computer Aided Modeling, Analysis and Design of Communication Links and
  Networks (CAMAD)}IEEE. ; 2010\string: 101--105.

\bibitem{p44}
Ker{\"a}nen A, K{\"a}rkk{\"a}inen T, Ott J. Simulating Mobility and DTNs with
  the ONE. {\it Journal of Communications} 2010\string; 5(2)\string: 92--105.

\bibitem{p45}
Daly EM, Haahr M. Social network analysis for routing in disconnected
  delay-tolerant manets. In:  {\it Proceedings of the 8th ACM international
  symposium on Mobile ad hoc networking and computing}ACM. ; 2007\string:
  32--40.

\end{thebibliography}
\end{document}